\documentclass[12pt]{article}

\input{epsf}
\usepackage{epsfig}
\usepackage{amssymb,amsmath}
\usepackage{amsfonts}
\usepackage{amsbsy}
\usepackage{empheq}
\usepackage{latexsym}
\usepackage{cite}

\newcommand{\beq}{\begin{equation}}
\newcommand{\eeq}{\end{equation}}
\newcommand{\bea}{\begin{eqnarray}}
\newcommand{\eea}{\end{eqnarray}}
\newcommand{\half}{\frac{1}{2}}

\newcommand{\la}{\langle}
\newcommand{\ra}{\rangle}

\newcommand{\EMB}{\mathbf}
\newcommand{\VEC}[1]{{\mathsf{#1}}}
\newcommand{\vt}[1]{{\mathbf{#1}}}
\newcommand{\uVEC}[1]{{\EMB{\hat{#1}}}}

\newcommand{\hz}{\EMB{\hat{z}}}
\newcommand{\hn}{\EMB{\hat{n}}}
\newcommand{\hx}{\EMB{\hat{x}}}
\newcommand{\hy}{\EMB{\hat{y}}}
\newcommand{\hza}{\EMB{\hat{z}_a}}
\newcommand{\hzb}{\EMB{\hat{z}_b}}
\newcommand{\hzc}{\EMB{\hat{z}_c}}
\newcommand{\hxa}{\EMB{\hat{x}_a}}
\newcommand{\hxb}{\EMB{\hat{x}_b}}
\newcommand{\hxc}{\EMB{\hat{x}_c}}
\newcommand{\hya}{\EMB{\hat{y}_a}}
\newcommand{\hyb}{\EMB{\hat{y}_b}}
\newcommand{\hyc}{\EMB{\hat{y}_c}}

\newcommand{\hxd}{\EMB{\hat{x}_d}}
\newcommand{\hxe}{\EMB{\hat{x}_e}}

\newcommand{\hv}{\EMB{\hat{v}}}
\newcommand{\hvf}[2]{\EMB{\hat{v}_{\left[\perp,#1,#2\right]}}}
\newcommand{\hvs}[2]{\EMB{\hat{v}_{\left[#1,\perp,#2\right]}}}
\newcommand{\hvth}[2]{\EMB{\hat{v}_{\left[#1,#2,\perp\right]}}}
\newcommand{\svf}[2]{v_{\left[\perp #1#2\right]}}
\newcommand{\svs}[2]{v_{\left[#1 \perp #2\right]}}
\newcommand{\svth}[2]{v_{\left[#1 #2 \perp\right]}}
\newcommand{\vv}{\vt v}
\newcommand{\vb}{\bar{v}}
\newcommand{\vbb}[1]{\bar{v}_{\textrm{\Rmnum{#1}}}}

\DeclareMathOperator{\Tr}{Tr}
\DeclareMathOperator{\diag}{diag}
\DeclareMathOperator{\sign}{sign}

\numberwithin{equation}{section}
\makeatletter

\newcommand{\Rmnum}[1]{\expandafter\@slowromancap\romannumeral #1@}
\makeatother

\begin{document}

\begin{titlepage}
\mbox{}
\vspace{25mm}
\begin{center}
{\Large\bf%
The Density Operators of Qubit Systems in the Multiparticle Spacetime Algebra
}\vspace*{15mm}

{\large Chih-Wei Wang} \
\vspace{1cm}

 \tt{\small freeform1111@gmail.com}

\end{center}

\vspace*{20mm}

\begin{abstract}

We provide a method to write down the density operator for any pure state of multi-qubit systems in the multiparticle spacetime algebra (MSTA) introduced by Doran, Gull, and Lasenby. Using the MSTA formulation, we analyze several aspects of quantum mechanics in a geometrical way including the Bell inequalities and the dynamics of two coupled qubits. Lastly, we provide a natural way to construct the local unitary invariants in the MSTA. Using these invariants, we analyze the space of the two-qubit and three-qubit pure states with local and non-local degrees of freedom separated.

\end{abstract}

\end{titlepage}

\newpage

\tableofcontents

%%%%%%%%%%%%%%%%%%%%%%%%
\section{Introduction}
%%%%%%%%%%%%%%%%%%%%%%%%

The conventional formulation of quantum mechanics is based on an abstract complex Hilbert space. However, it is undeniable that some geometric pictures give us great insights into some quantum mechanics or quantum information problems. For example, it is very helpful to some problems or algorithms by picturing a pure state of a qubit as a unit vector of the Bloch sphere. However, these geometric pictures are scarce and limited. On the other hand, there is a new approach to quantum mechanics developed by Hestenes\cite{hestenes1966space,hestenes1971vectors,hestenes1979spin} based on the geometric (Clifford) algebra. This new approach formulates quantum mechanics in a more geometrical way and makes the connection between quantum mechanics and the space-time more transparent. In this paper, we will follow this approach and try to develop a geometrical method to analyze qubit systems.

To describe multi-particle systems in the framework of geometric algebra (GA), we will use the \emph{multiparticle spacetime algebra} (MSTA) introduced by Doran, Gull, and Lasenby \cite{doran1993states,doran1996spacetime}. Furthermore, in order to deal with pure and mixed states alike, we will focus on the density operators. There are in fact two different representations for the density operators in the MSTA depending on how we deal with the imaginary structures from the multiple qubits\cite{havel2003density}. The first representation is achieved by correlating the pseudoscalars of different qubits and the result is basically a direct transliteration from the usual Hermitian density matrix into the MSTA. The alternated one is done by correlating some chosen bivectors. The intention of the second representation is to generalize Hestenes' operator view of a spinor to the multi-particle regime. However, it will result in the non-hermitian density operator. We will follow the first approach but we will show that it does not need to lose the appeals of the original view by using the projectors accordingly.

The main purpose of this paper is to show the usefulness of the MSTA in describing the quantum system in a geometrical way. Therefore, instead of formal completeness, we will focus on demonstrating several simple but nontrivial examples including the Bell inequality and the dynamics of two coupled qubits. Furthermore, one of the most important features that distinguish quantum and classical systems is entanglement. In order to study the different types of entanglement, we should use several quantities which will not be changed under the local unitary transformations to parametrize the non-local part of the state space. Since the effects of the local unitary transformations are quite transparent in the MSTA, there is a natural way to define the local unitary invariants. These invariants can be used to analyze the non-local part of the state space and study the entanglement.

We assume the readers are familiar with geometric algebra (a good introduction can be found in \cite{doran2003geometric,baylis2012clifford}) and only briefly review the multiparticle spacetime algebra in section 2. We will show how to write down the density operator for any pure state of multi-qubit systems and study its structure in section 3. We will also show how to study the Bell inequality and the dynamic of two couple qubits in a geometrical way in section 4 and 5.  In section 6, we show how to construct a set of the invariants for the two-qubit and three-qubit pure states in the MSTA. Using these invariants, we analyze their state spaces with the local and non-local degrees of freedom separated. In section 7, we summarize and point out several directions we can proceed in the future.

%%%%%%%%%%%%%%%%%%%%%%%%%%%%%%%%%%%%%%%%%%%%%%%%%%%%%
\section{Multiparticle spacetime algebra}

Spacetime algebra introduced by Hestenes\cite{hestenes1966space} is basically the real Clifford algebra of Minkowski spacetime denoted by $\mathcal{G}(1,3)$.  The algebra is generated by a basis $\{\gamma_\mu\}$ which satisfy:
\beq
\gamma_{\mu}\,\gamma_{\nu} + \gamma_{\nu}\,\gamma_{\mu}  ~=~ 2\,\eta_{\mu\nu}\,,~\quad~\mu.\nu=0,\cdots,3,
\eeq
where $\eta_{\mu\nu}$ is the metric tensor:
\beq
\eta_{\mu\nu} = \diag(+1,-1,-1,-1)\,.
\eeq

The Pauli algebra, $\mathcal{G}(3)$, is spatial part of the spacetime algebra and is generated by:
\beq
\hx = \gamma_1\gamma_0\,, ~\quad~ \hy = \gamma_2\gamma_0\,, ~\quad~\hz = \gamma_3\gamma_0\,.
\eeq
The three basis vectors are unitary and anticommute with each other:
\beq
\hx\,\hy=-\hy\,\hx=I\hz\,, ~\quad~ \hx\,\hz=-\hz\,\hx=-I\hy\,, ~\quad~\hy\,\hz=-\hz\,\hy=I\hx\,.
\eeq
where $I$ is the pseudoscalar:
\beq
I ~\equiv~ \gamma_0\gamma_1\gamma_2\gamma_3 ~=~ \hx\,\hy\,\hz\,.
\eeq

The $N$-particle MSTA, $\mathcal{G}(N,3N)$, is the algebra associated with $N$ copies of Minkowski spacetime which represent the configuration space of the $N$-particle system. It can be generated by a basis $\{\gamma_\mu^a\}$ which satisfy:
\beq
\gamma_{\mu}^a\,\gamma_{\nu}^b + \gamma_{\nu}^b\,\gamma_{\mu}^a  ~=~ 2\,\eta_{\mu\nu}\delta^{ab}\,,~\quad~a,b=0,\cdots,N,
\eeq
where the superscripts label the particle spaces. Inside the $N$-particle MSTA, there are N copies of the Pauli algebra generated by:
\beq
\hx_a = \gamma_1^a\gamma_0^a\,, ~\quad~ \hy_a = \gamma_2^a\gamma_0^a\,, ~\quad~\hz_a = \gamma_3^a\gamma_0^a\,, ~\quad~ a=0,\cdots,N\,,
\eeq
where we use the subscripts to denote the particle spaces for the Pauli algebra. It can be verified that any two vectors from different particle spaces commute. Therefore, the subalgebra generated by these vectors are isomorphic to the tensor product of $N$ copies of the Pauli algebra $\mathcal{P}(N)$: $\mathcal{G}(3)\otimes\cdots\otimes\mathcal{G}(3)$. This is the algebra we are going to use in this paper since we focus on non-relativistic quantum mechanics. Furthermore, in this work, we will use the following fonts to denote the different elements of MSTA: `$a$' as a scalar, `$\mathbf{a}$' as a general vector in $\mathcal{G}(3)$,  `$\mathbf{\hat{a}}$' as a unit vector in $\mathcal{G}(3)$ and `$\mathsf{A}$' as a general multivector that may contain the product of several vectors from different $\mathcal{G}(3)$.

Before we can use $\mathcal{P}(N)$ to describe multi-particle systems, we need to deal with the pseudoscalars from different particle spaces. To make the complex structures comparable to quantum mechanics, the pseudoscalar correlator was introduced\cite{havel2000geometric,havel2003density}:
\beq
C=\prod_{a=2}^N\,\frac{1}{2}\,(1-I_1 I_a)\,.
\eeq
This correlator reduces multiple pseudoscalars to a single correlated one:
\beq
\label{cori}
\iota \equiv I_1C =\cdots =I_NC\,.
\eeq
This correlated pseudoscalar will play the role of the imaginary unit $i$ of traditional quantum mechanics. Since the only effect of this correlator is to identify the pseudoscalars from different particle spaces, we will omit it and simply replace any encountered pseudoscalar with the correlated one.

An important operation we may need is the reverse operation. This operation reverse the order of the vectors for each particle space and is denoted by the dagger symbol. For example:
\beq
(\vt m_a\, \vt n_a\, \vt p_b\, \vt q_b)^{\dag} ~=~ \vt n_a\, \vt m_a\, \vt q_b\, \vt p_b\,.
\eeq
This operation is equivalent to Hermitian conjugate in the matrix formulation.

%reverse operation
%geometric product

%%%%%%%%%%%%%%%%%%%%%%%%%
\section{Density operators in the MSTA}

In this section, we start from the Bloch sphere representation of a single qubit. By using the GA, we can write down the density operator for any state of a qubit similarly. But, the formulation and computation can be done in a coordinate-free way. Then we further generalize it to two-qubit and three-qubit systems in the MSTA. Additionally, we show that any unitary transformation can be realized as a series of rotations in some projective spaces. Finally, we study the structure of the density operator for any pure state of multi-qubit systems.

%%%%%%%%%%%%%%%%%%%%%%%%%
\subsection{Single qubit}

It is well known that the density operator of a single qubit can be expressed in the following form:
\beq
\rho ~=~ \frac{I ~+~ n_x\,\sigma_x ~+~ n_y\,\sigma_y ~+~ n_z\,\sigma_z}{2}\,,
\eeq
where $\sigma_x$, $\sigma_y$ and $\sigma_z$ are Pauli matrices. The scalars $n_x$, $n_y$ and $n_z$ are the components of the Bloch vector $\vec{n}$ and they satisfy $n_x^2+n_y^2+n_z^2 \leq 1$. Therefore, the density operator of a qubit is parametrized by a vector inside of a unit ball, the Bloch sphere. When the state is pure, the vector is unitary and the state is on the sphere.

On the other hand, the similar form can be found in the GA formulation. In \cite{hestenes1966space}, it was pointed out that spinors have the same geometrical meaning as rotors by extracting the projection factor:
\beq
\Psi ~=~ \psi\,\left(\frac{1+\hz}{2} \right)\,,
\eeq
where $\psi$ is restricted to the even subalgebra of the Pauli algebra and can be interpreted as a rotor if it is normalized. Therefore, the density operator of a pure state is:
\beq
\rho  ~=~ \Psi\Psi^{\dag} ~=~ \psi \left(\frac{1+\hz}{2} \right) \psi^{\dag} ~=~ \frac{1+\hn}{2}\,,
\eeq
where $\hn = \psi\hz\psi^{\dag}$. Therefore, a spinor can be considered as a rotor which rotates some reference state (which is conventionally defined by the z-axis) to the actual state. For a mixed state, the vector becomes non-unitary: $\hn \rightarrow \vt n$ with $\vt n^2<1$. Comparing to the usual matrix formulation, this coordinate-free approach is one of the advantages of the GA/MSTA formulation.

In order to compute the expectation values of the physical observables, we need to translate the trace operation to the GA/MSTA formulation. It can be shown that the trace is simply proportional to the scalar part of the corresponding expression: 
\beq
\label{TrinMSTA}
\Tr(\rho) ~\rightarrow~ \mathcal{N} \la \rho \ra\,,
\eeq
where $\mathcal{N}$ is the number of the states that have been traced (e.g., $\mathcal{N}=2^n$ for an n-qubit system). The operation $\la\_\ra$ extract the scalar (i.e., grade zero) part. 

To show how to do the computation in this formulation, we can consider some projective measurements described by ${E_\pm}= (1 \pm \EMB{\hat{s}})/2$, the probabilities of the measurements on some generic pure state $\rho=(1 + \hn)/2$ are:
\beq
\Tr(E_\pm\,\rho) ~\rightarrow~ 2 \langle E_\pm \rho \rangle ~=~ \frac{1 \pm \EMB{\hat{s}}\cdot\hn}{2} ~=~ \frac{1 \pm \cos\theta }{2}\,,
\eeq 
where $\theta$ is the angle between the two vectors $\hn$ and $\EMB{\hat{s}}$.

%%%%%%%%%%%%%%%%%%%%%%%%%%%%%%
\subsection{Two qubits}

It is quite straightforward to construct the density operator for any product state of multi-qubit systems. For example, any pure product state of the two qubits labeled with `$a$' and `$b$' can be expressed as:
\beq
\rho_{ab} ~=~ \left(\frac{1+\uVEC m_a}{2}\right)\left(\frac{1+\uVEC n_b}{2}\right)\,,
\eeq
where $\uVEC m_a$ and $\uVEC n_b$ are some unit vectors of the qubits $a$ and $b$ respectively. However, the two qubits can become entangled through the interaction which is one of the most important features of quantum mechanics. Therefore, we need to find a way to write down the density operators for the entangled states as well. 

Based on the Schmidt decomposition, any pure state of two qubits can be decomposed to the superposition of two orthogonal product sates. From this perspective, an entangled state can also be regarded as some kind of two-level system just like a single qubit. Therefore, we should be able to use a sphere to describe the space of the entangled states in a similar way as the Bloch sphere.  A similar idea, the \emph{entanglement sphere}, and a similar parametrization have been proposed in \cite{wie2014bloch,wharton2016natural}.

To identify the rotors that rotate this sphere, we need to specify the three basis \emph{(multi)-vectors} of the sphere's space. To simplify the problem for now, we will assume the two product states in the Schmidt decomposition are:
\beq
\{00\} \equiv \left(\frac{1+\uVEC z_a}{2}\right)\left(\frac{1+\uVEC z_b}{2}\right)\,, ~\quad~ 
\{11\} \equiv \left(\frac{1-\uVEC z_a}{2}\right)\left(\frac{1-\uVEC z_b}{2}\right)\,.
\eeq
Since they are pure states, they satisfy: $\{00\}^2=\{00\}$ and $\{11\}^2=\{11\}$.
The projector that project a general pure state to the space of the superposition of these two states is:
\beq
\VEC P_{\{00,11\}} ~\equiv~ \{00\} + \{11\} ~=~ \frac{1+ \uVEC z_a \uVEC z_b}{2}\,.
\eeq
Since the two product states are orthogonal, it is clearly that this operator satisfy the requirement of a projector: $\VEC P^2_{\{00,11\}} =\VEC P_{\{00,11\}} $. The three basis vectors in this projector space can be defined by:
\bea
\VEC Z_{\{00,11\}} &\equiv& \hz_a\, \VEC P_{\{00,11\}} ~=~ \hz_b\, \VEC P_{\{00,11\}} ~=~ \frac{1}{2}\,(\hz_a+\hz_b)\,,\\
\VEC X_{\{00,11\}} &\equiv& \hx_a\, \hx_b\, \VEC P_{\{00,11\}} ~=~ -\hy_a\,\hy_b\, \VEC P_{\{00,11\}} ~=~ \frac{1}{2}\,(\hx_a\,\hx_b-\hy_a\,\hy_b)\,, \\
\VEC Y_{\{00,11\}} &\equiv& \hx_a\, \hy_b\, \VEC P_{\{00,11\}} ~=~ \hy_a\,\hx_b\, \VEC P_{\{00,11\}} ~=~ \frac{1}{2}\,(\hx_a\,\hy_b + \hy_a\,\hx_b)\,,
\eea

Even though they are in fact multi-vectors in MSTA, they commute with $\VEC P_{\{00,11\}}$ and satisfy the Pauli algebra $\mathcal{G}(3)$ in the projector space:
\bea
&\VEC Z_{\{00,11\}}^2 = \VEC X_{\{00,11\}}^2 = \VEC Y_{\{00,11\}}^2 = \VEC P_{\{00,11\}}\,,~\quad~ 
\VEC Z_{\{00,11\}}\, \VEC X_{\{00,11\}} = - \VEC X_{\{00,11\}}\,\VEC Z_{\{00,11\}}\,, \\
&\VEC Z_{\{00,11\}}\, \VEC Y_{\{00,11\}} = - \VEC Y_{\{00,11\}}\,\VEC Z_{\{00,11\}}\,, ~\quad~
\VEC X_{\{00,11\}}\, \VEC Y_{\{00,11\}} = - \VEC Y_{\{00,11\}}\,\VEC X_{\{00,11\}}\,, \\
&\VEC X_{\{00,11\}}\,\VEC Y_{\{00,11\}}\,\VEC Z_{\{00,11\}}= \iota\, \VEC P_{\{00,11\}}\,,
\eea
where $\iota$ is the correlated pseudoscalar introduced in \eqref{cori}. Therefore we can treat them as the three basis vectors associated with the projector space. 

The \emph{entanglement} sphere is a unit sphere in the space expanded by the three basis vectors. The rotor that rotate the sphere can be constructed as the usual. For example, the rotor that describe the rotation around $\VEC Y_{\{00,11\}}$ is:
\bea
\label{rotor2q}
e^{-\iota \VEC Y_{\{00,11\}} \theta/2} &=& 1 + (\cos\frac{\theta}{2}-1)\,\VEC P_{\{00,11\}} +\sin\frac{\theta}{2}\,\iota \VEC Y_{\{00,11\}}\,, \nonumber \\
&=& (1- \VEC P_{\{00,11\}}) + e^{-\iota \hx_a\hy_b \theta/2}\, \VEC P_{\{00,11\}}\,,
\eea
where $(1- \VEC P_{\{00,11\}})$ is the orthogonal projector:
\beq
1- \VEC P_{\{00,11\}} ~=~ \VEC P_{\{01,10\}}  ~\equiv~  \{01\} + \{10\} ~=~  \frac{1 - \uVEC z_a \uVEC z_b}{2}\,.
\eeq
Therefore, the rotor in \eqref{rotor2q} only perform the rotation in the projector space $\VEC P_{\{00,11\}}$ while do nothing in the orthogonal space. We can also define the three basis vectors for the $\VEC P_{\{01,10\}}$:
\bea
&\VEC Z_{\{01,10\}}  ~\equiv~ \hz_a\, \VEC P_{\{01,10\}} ~=~ -\hz_b\, \VEC P_{\{01,10\}} ~=~ \frac{1}{2}\,(\hz_a-\hz_b)\,,\\
&\VEC X_{\{01,10\}}  ~\equiv~ \hx_a \hx_b\, \VEC P_{\{01,10\}} ~=~ \hy_a\hy_b\, \VEC P_{\{01,10\}} ~=~ \frac{1}{2}\,(\hx_a\hx_b + \hy_a\hy_b)\,, \\
&\VEC Y_{\{01,10\}}  ~\equiv~ \hy_a \hx_b\, \VEC P_{\{01,10\}} ~=~ -\hx_a\hy_b\, \VEC P_{\{01,10\}} ~=~ \frac{1}{2}\,(\hy_a\hx_b - \hx_a\hy_b )\,.
\eea
The rotor in this projector space can be constructed similarly and it only rotate the space $\VEC P_{\{01,10\}}$ while do nothing in $\VEC P_{\{00,11\}}$.

Any point on the sphere represents a pure state in the projector space. To construct the density operator for any state in $\VEC P_{\{00,11\}}$, we can use the product state $\{00\}$ as the reference state as usual. This state is located at the north pole of the sphere:
\beq
\{00\} ~=~ \frac{1}{2}\,(\VEC P_{\{00,11\}} + \VEC Z_{\{00,11\}})\,.
\eeq
Then any other state can be obtained by applying some rotor on $\{00\}$:
\beq
\label{rhos}
\rho_{\VEC S} ~=~ e^{-\iota \VEC Z_{\{00,11\}}\,\phi/2}\,e^{-\iota \VEC Y_{\{00,11\}}\,\theta/2}\,\{00\}\,e^{\iota \VEC Y_{\{00,11\}},\theta/2}\,e^{\iota \VEC Z_{\{00,11\}}\,\phi/2}
~=~  \frac{1}{2}\,(\VEC P_{\{00,11\}} + \VEC S_{\{00,11\}})\,,
\eeq
where $\VEC S_{\{00,11\}}$ is some unit vector on the sphere:
\beq
\VEC S_{\{00,11\}} ~=~ \cos\theta\, \VEC Z_{\{00,11\}}~+~ \sin\theta\cos\phi\, \VEC X_{\{00,11\}} ~+~ \sin\theta\sin\phi\, \VEC Y_{\{00,11\}}\,.
\eeq
The angle $\theta$ is directly related to the entanglement between two qubits which will be explained later. Combining the reference state, the spinor of the above state is:
\beq
\Psi_{\VEC S} ~=~ \psi_{\VEC S}\,\{00\} ~=~  e^{-\iota \VEC Z_{\{00,11\}}\,\phi/2}\,e^{-\iota \VEC Y_{\{00,11\}}\,\theta/2}\,\{00\}\,.
\eeq
Furthermore, using the unitarity of $\VEC S_{\{00,11\}}$ (i.e., $\VEC S_{\{00,11\}}^2 = \VEC P_{\{00,11\}}$), it can be easily verified that the density operator satisfy the pure state condition: $\rho_{\VEC S}^2=\rho_{\VEC S}$.

To find the density operator for a general two-qubit pure state, we simply need to use more general orthogonal product states for the projector. For example, we can consider the following two states:
\beq
\{++\} =\left(\frac{1+\uVEC m_a}{2}\right)\left(\frac{1+\uVEC n_b}{2}\right)\,,~\quad~ \{--\} =\left(\frac{1-\uVEC m_a}{2}\right)\left(\frac{1-\uVEC n_b}{2}\right)\,,
\eeq
and the new projector will be:
\beq
\VEC P_{\{++,--\}} ~=~ \{++\} + \{--\} ~=~ \frac{1+\uVEC m_a\,\uVEC n_b}{2}\,.
\eeq
We can follow the similar procedure to construct the new basis vectors for the sphere's space and the density operator. However, we will find out that the result is related to the density operator in the previous projector by the local rotations of the qubits $a$ and $b$ (i.e., the local unitary transformation). The rotations are described by some local rotors that connect the product states in the two projectors:
\beq
R_a\, (\hx_a,\hy_a,\hz_a)\,R_a^\dag  = (\uVEC m_{a\dashv}\,,\uVEC m_{a\vdash}\,,\uVEC m_a)\,, ~\quad~
R_b\, (\hx_b,\hy_b,\hz_b)\,R_a^\dag  =  (\uVEC n_{b\dashv}\,,\uVEC n_{b\vdash}\,,\uVEC n_b)\,,
\eeq
where $ (\uVEC m_{a\dashv}\,,\uVEC m_{a\vdash}\,,\uVEC m_a)$ and $(\uVEC n_{b\dashv}\,,\uVEC n_{b\vdash}\,,\uVEC n_b)$ form the new bases for the space of the qubit $a$ and $b$ respectively.
Then the density operator of a general pure state can be written as:
\beq
\rho'_{\VEC S} ~=~ \frac{1}{2}\,(\VEC P_{\{++,--\}} + \VEC S_{\{++,--\}})\ ~=~R_a\,R_b\,\rho_{\VEC S}\,R^\dag_b\,R^\dag_b\,.
\eeq
A general spinor of two qubits can be written in this form:
\beq
\Psi_{\VEC S'} ~=~ R_a\,R_b\,\Psi_{\VEC S} ~=~ R_a\,R_b\,e^{-\iota \VEC Z_{\{00,11\}}\,\phi/2}\,e^{-\iota \VEC Y_{\{00,11\}}\,\theta/2}\,\{00\}\,.
\eeq

As we have explained in \eqref{TrinMSTA}, the trace operation in the MSTA is done by simply extracting the scalar part while dumping anything else. The expectation value of any physical observable for a general two-qubit pure state can be computed by:
\beq
\Tr(\mathcal{O}\,\rho_{\VEC S'}) ~\rightarrow~ 4\,\la \mathcal{O}\,\rho_{\VEC S'} \ra\,,
\eeq
where $\mathcal{O}$ should be expressed in the MSTA. Furthermore, since any mixed state can be expressed as a probability-weighted sum of several pure states, we should be able to use this formulation to compute any physical quantity for a general two-qubit system. 

In some situations, we may need to do the partial trace operations. This can be achieved by simply dumping any term contains the vectors of the qubits which have been traced out. For example, the reduced density operator of the first qubit $\rho_a$ can be obtained by tracing out the qubit $b$ from the $\rho_{\VEC S}$ in \eqref{rhos}:
\beq
\label{rhos_trb}
\Tr_b(\rho_{\VEC S}) ~\rightarrow~ 2\,\left\la \rho_{\VEC S} \right\ra_b ~=~ \frac{1}{2}\,(1+\cos\theta\, \hza) ~=~ \rho_a\,,
\eeq
where the coefficient, $2$, is the number of the states that has been traced out. This operation $\la\_\ra_b$ simply drop the terms that contains any vector of the qubit $b$. 

From the above expression, we can see that the $\rho_a$ is no longer a pure state if the angle $\theta$ is not zero or $\pi$ because the vector is no longer unitary. The length of this vector is related to the von Neumann entropy of the $\rho_a$ which can be used to measure the entanglement between two qubits:
\beq
S ~=~ - \Tr\rho_a \log_2 \rho_a  ~=~ - \Tr\rho_b \log_2 \rho_b\,.
\eeq
To compute the Von Neumann entropy in the MSTA for this simple case, it may be more convenient to take the limit of the Renyi entropy:
\beq
S~=~ \lim_{\alpha \rightarrow 1} \frac{1}{1-\alpha}\,\log_2\,(\Tr(\rho_a^\alpha)) ~\rightarrow~ \lim_{\alpha \rightarrow 1} \frac{1}{1-\alpha}\,\log_2\,(2\,\la \rho_a^\alpha \ra)
\eeq
where $\la\rho_a^{\alpha}\ra$ is:
\beq
\la \rho_a^{\alpha} \ra =\frac{1}{2^{\alpha}}\,\left\la(1+\cos\theta\,\hza)^\alpha\right\ra = \frac{1}{2^{\alpha+1}}\,\left( (1+\cos\theta)^{\alpha}+(1-\cos\theta)^{\alpha}\right)\,.
\eeq
So the von Neumann entropy is:
\beq
\label{S_tq}
S~=~ -\left(\frac{1-\cos\theta}{2}\right)\log_2\,\left(\frac{1-\cos\theta}{2}\right) - \left(\frac{1+\cos\theta}{2}\right)\log_2\,\left(\frac{1+\cos\theta}{2}\right)\,.
\eeq
Therefore the angle $\theta$ determine the entanglement between two qubits and the entropy reach the maximum $S=1$ when $\theta$ is equal to $\pi/2$. At that point, the $\rho_a$ becomes completely mixed and its vector part vanishes. That is when the two qubits become maximally-entangled.

%%%%%%%%%%%%%%%%%%%%%%%%%%%%%%%%%%%%%%%
\subsection{Generalized spheres and three-qubit examples}

Previously, we have identified the spheres of the entangled states in the projector $\VEC P_{\{00,11\}}$ and $\VEC P_{\{01,10\}}$. In fact, any projector built from two orthogonal product states will have the similar sphere. For example, we can consider the following projector:
\beq
\VEC P_{\{00,01\}} = \{00\} + \{01\} = \frac{1+\hza}{2}\,.
\eeq
The three basis vectors associated with this projector will be:
\beq
\VEC Z = \hzb \frac{1+\hza}{2}\,,~\quad~ \VEC X = \hxb \frac{1+\hza}{2}\,, ~\quad~ \VEC Y = \hyb \frac{1+\hza}{2}\,.
\eeq
The sphere is actually just the Bloch sphere of the qubit $b$ times the projector $(1+\hza)/2$. The rotors related to this sphere are just the \emph{conditional} local rotations acting on the qubit $b$ (i.e., they only rotate the qubit $b$ when the qubit $a$ is `$0$'). Out of the six projectors coming from the different combinations of two product states, four of them are basically just some conditional local rotations like the above example. The only two exceptions are $\VEC P_{\{00,11\}}$ and $\VEC P_{\{01,10\}}$ which have been introduced previously.

This procedure can be generalized to any projector built from two orthogonal product states\footnote{It may be possible to generalize to any two orthogonal pure states. But, it is unclear so far how to identify the three basis vectors in general.} of the multi-qubit systems. For example, we can consider the following projector of a three-qubit system:
\beq
\VEC P_{\{000,111\}} \equiv \{000\} + \{111\} = \frac{1}{4}(1+\hza\hzb+\hza\hzc+\hzb\hzc)\,.
\eeq
We will demonstrate the procedure to find the three basis vectors for the sphere associated with this projector. In the following, we will denote them by $\VEC X$, $\VEC Y$, and $\VEC Z$ without any super- or sub-script and it should be understood that they are in the projector defined above. 

At first, we need to decide the ordering of the two states in order to define the vector $\VEC Z$. That is whether we set $\VEC Z$ equal to $\{000\}-\{111\}$ or $\{111\}-\{000\}$. This choice will affect the handedness of the azimuthal angle but other than that there should not have any physical significance. We will take the first choice:
\beq
\label{Zabc_def}
\VEC Z ~\equiv~ \{000\}-\{111\} ~=~ \frac{1}{4}(\hza + \hzb + \hzc + \hza\hzb\hzc)\,.
\eeq
The required properties of the $\VEC X$ are:
\beq
\VEC X\, \VEC P_{\{000,111\}}=\VEC P_{\{000,111\}}\, \VEC X = \VEC X\,, ~\quad~ \VEC X^2 = \VEC P_{\{000,111\}}\,, ~\quad \VEC X\, \VEC Z = -\VEC Z\,\VEC X\,.
\eeq
To find the $\VEC X$, consider the following fact:
\beq
\hxa\hxb\hxc\,\{000\}= \{111\}\, \hxa\hxb\hxc\,.
\eeq
Note that any vector orthogonal to $\hz$ will have the similar effect as the $\hx$. This implies that $\VEC X$ should contain the product of several unit vectors of \emph{only} those qubits with the opposite signs in the two states of the projector and each of these unit vectors should be orthogonal to the spin of the corresponding qubit. This product will be square to one, commute with the projector and anti-commute with $\VEC Z$. In order to maintain the simpler relation with the density operator in the conventional matrix formulation, we will choose the following definition for the $\VEC X$:
\beq
\VEC X ~\equiv~ \hxa\hxb\hxc\,\VEC P_{\{000,111\}} = \frac{1}{4}(\hxa\hxb\hxc-\hxa\hyb\hyc-\hya\hxb\hyc-\hya\hyb\hxc) \,.
\eeq

Because of the presence of the projector, the following definitions are also true:
\beq
\label{Xabc_def}
 \VEC X ~=~ -\hxa\hyb\hyc\,\VEC P_{\{000,111\}} =-\hya\hxb\hyc\,\VEC P_{\{000,111\}} =-\hya\hyb\hxc\,\VEC P_{\{000,111\}} \,.
\eeq
In fact, any simultaneous azimuthal rotations of two unit vectors in the opposite ways will not change $\VEC X$. For example,
\beq
\label{3q_oprot}
\hxa\,(\cos\phi\, \hxb + \sin\phi\, \hyb)\,(\cos\phi\, \hxc - \sin\phi\, \hyc)\,\VEC P_{\{000,111\}} ~=~ \VEC X\,.
\eeq
Therefore, although the vector $\VEC X$ seems to have three degrees of freedom coming from the azimuthal angles of the three qubits, there is actually only one azimuthal angle associated to this projector space.

The remaining vector $\VEC Y$ can be found by using the fact: $\VEC X \VEC Y \VEC Z = \iota \VEC P_{\{000,111\}}$,
\bea
\VEC Y &=& \iota\, \VEC X\, \VEC Z = \iota\,\hxa\hxb\hxc\,(\{000\}-\{111\}) \nonumber \\
&=& \hxa\hxb\hyc\,\VEC P_{\{000,111\}} = \hxa\hyb\hxc\,\VEC P_{\{000,111\}} = \hya\hxb\hxc\,\VEC P_{\{000,111\}}\nonumber \\
&=& \frac{1}{4}(\hxa\hxb\hyc-\hya\hyb\hyc+\hya\hxb\hxc+\hxa\hyb\hxc)\,.
\eea
As we have expected, we can rotate $\VEC X$ to $\VEC Y$ by the local rotation of one of the qubits. Notice that in the eight combinations: $\hxa\hxb\hxc\,,\hxa\hxb\hyc\cdots$, those with even number of $\EMB{ \hat{y}}$ are all projected to $\VEC X$ or $-\VEC X$ and the ones with odd number of $\EMB{ \hat{y}}$ will be projected to $\VEC Y$ or $-\VEC Y$. From the definition of the $\VEC X$ and $\VEC Y$, we see that the rotating of any small $\hx$ with the azimuthal angle $\phi$ will also rotate the $\VEC X$ with the same angle. However, if we pick $\VEC Z=\{111\}-\{000\}$, $\VEC Y$ will be flipped. In that case, the rotation angle of the $\VEC X$ will become $-\phi$. This is simply because the definition of $\VEC Z$ and the azimuthal angle of the projector sphere have changed but the effect of the rotating of $\hx$ on $\VEC X$ is still the same. 

To save us from the confusion related to the effects of the local axial rotations of qubits on the azimuthal angle of the $\VEC X$, we will point out the following rule. For a single qubit, the definition of $\hz$ is always $\{0\}-\{1\}$ which fix the ordering for any single qubit. If the ordering in the definition of the $\VEC Z$ agrees with the ordering of an individual qubit then the induced change of the azimuthal angle of the $\VEC X$ will be the same as the rotation angle of that qubit. If not, the sign of the change will be opposite. For example, for the projector, $\VEC P=\{01\} + \{10\}$ and $\VEC Z = \{01\}-\{10\}$, the rotating of the $\hxa$ by $\phi$ will rotate the $\VEC X$ by the same angle while the rotating of the $\hxb$ will have the opposite effect. This means if we rotate $\hxa$ and $\hxb$ with the \emph{same} angles, the $\VEC X$ remains unchanged in this case.

The three basis vectors, $\VEC Z$, $\VEC X$ and $\VEC Y$ satisfy the required properties for a basis associated with the projector. Therefore, one can build any rotor that rotates the unit sphere just like the two-qubit case. Any unitary transformation in this projector space can be regarded as some rotation generated by a rotor. Any pure state in the projector space can be identified as a point on the sphere. To illustrate this explicitly, we can consider the unitary transformation that takes $|000\ra$ to $\alpha |000\ra + \beta |111\ra$ with the coefficients normalized. The global phase of the state has no physical meaning but the phase difference of the coefficients will be related to the azimuthal angle of the sphere. If we follow the definitions of $\VEC Z$ and $\VEC X$ in \eqref{Zabc_def} and \eqref{Xabc_def}, the points on the full circle along the $\VEC X$ direction represent the real states (i.e., the state with both the coefficients real after removing the global phase). The azimuthal angle of a point on the sphere should be related to the coefficients of the  corresponding state by:
\beq
\phi =  \arg\left(\frac{\beta}{\alpha}\right)\,.
\eeq

The above unitary transformation corresponds to the following rotation:
\begin{align}
&e^{-i \VEC X_{\phi+\frac{\pi}{2}} \frac{\theta}{2}}\,\{000\}\,e^{i \VEC X_{\phi+\frac{\pi}{2}} \frac{\theta}{2}} \nonumber\\
&=\left(\cos\frac{\theta}{2} -i\sin\frac{\theta}{2}\uVEC x_{a,\phi+\pi/2} \hxb\hxc\right)\{000\}\left(\cos\frac{\theta}{2}+i\sin\frac{\theta}{2}\uVEC x_{a,\phi+\pi/2} \hxb\hyc\right) \nonumber \\
\label{3q_pqf}
&=\cos^2\frac{\theta}{2}\, \{000\} + \sin^2\frac{\theta}{2}\,\{111\}+\sin\frac{\theta}{2}\cos\frac{\theta}{2}\uVEC x_{a,\phi}\hxb\hxc\,(\{000\}+\{111\}) \\
\label{3q_ZX}
&=\half\left(\VEC P_{\{000,111\}} + \cos\theta\, \VEC Z + \sin\theta\, \VEC X_\phi \right)\,,
\end{align}
where the rotation angle $\theta$ are related to the absolute values of the coefficients (i.e., $|\alpha|$ and $|\beta|$) and the $\hx_{a,\phi}$ and the $\VEC X_{\phi}$ are defined by:
\beq
\uVEC x_{a,\phi} \equiv \cos\phi\,\hxa + \sin\phi\,\hya. ~\quad~ \VEC X_{\phi} \equiv \cos\phi\,\VEC X + \sin\phi\,\VEC Y\,.
\eeq
The last form in \eqref{3q_ZX} is what we expect for such kind of the rotation. However, the expression in \eqref{3q_pqf} is also quite interesting since it has this kind of the structure: $p\{000\} + q\{111\} + \sqrt{p q}\, \VEC X_\phi$  with $p+q=1$. The values $p$ and $q$ are the probabilities associated to the two product states. They are related to $\alpha$ and $\beta$ by:
\beq
p=|\alpha|^2\,, ~\quad~ q=|\beta|^2\,.
\eeq
Therefore, we can say that the first two terms represent the diagonal components of the density operator in the conventional matrix formulation, while the last term, $\VEC X_\phi$, represent the off-diagonal components. Notice that the coefficient of the $\VEC X_{\phi}$ is completely determined by the two probabilities $p$ and $q$. This constraint can be traced back to the pure state condition and will be explained later.

For the rotation with $\phi=0$ and $\theta=\pi/2$, the result is the well-known GHZ state: $\frac{1}{\sqrt{2}}\left( |000\ra  + |111\ra \right)$. The corresponding density operator in the MSTA is:
\bea
\label{do_GHZ}
\rho_{GHZ} &=& \frac{1}{8}\,(1 + \hza\hzb + \hza\hzc + \hzb\hzc)\,(1+ \hxa\hxb\hxc ) \nonumber \\
&=&\frac{1}{8}\,\left(1 + \hza\hzb + \hza\hzc + \hzb\hzc  + \hxa\hxb\hxc - \hxa\hyb\hyc - \hya\hxb\hyc -\hya\hyb\hxc \right)\,.
\eea
This state has the maximum three-way entanglement of three qubits and it will be discussed further later.

Using the above procedure, we can basically construct the density operator in the MSTA for any superposition of two orthogonal product states. For the superposition of more than two states, we can apply several rotations sequently. For example, for the unitary transformation: $|000\ra$ to $\alpha |000\ra + \beta |011\ra + \gamma |110\ra$, we can break it into two sequential rotations in the projector spaces $\VEC P_{\{000,011\}}$ and $\VEC P_{\{011,110\}}$. The corresponding sequential rotors are:
\beq
e^{-i \VEC X_{2,\phi_2}  \frac{\theta_2}{2}}\,e^{-i \VEC X_{1,\phi_1}  \frac{\theta_1}{2}}\,,
\eeq
where $\VEC X_{1,\phi_1}$ and $\VEC X_{2,\phi_2}$ are associated to $\VEC P_{\{000,011\}}$ and $\VEC P_{\{011,110\}}$ respectively. Note that $\VEC X_1$ and $\VEC X_2$ do not commute. The azimuthal angles $\phi_1$ and $\phi_2$ are related to the phase differences of the coefficients, $\alpha$, $\beta$ and $\gamma$. The rotation angles, $\theta_1$ and $\theta_2$ are determined by the absolute values of the coefficients. Applying the rotors to $\{000\}$, the result will have the following structure:
\beq
\label{3q_sf}
p\,\{000\} + q\,\{011\} + r\, \{110\} + \sqrt{p\,q}\, \VEC X_{1,\psi_1} + \sqrt{q\,r}\, \VEC X_{2,\psi_2} + \sqrt{p\,r}\, \VEC X_{3,\psi_3}\,,
\eeq
where $p$, $q$ and $r$ are:
\beq
p=|\alpha|^2=\cos^2\frac{\theta_1}{2}\,,~~~  q=|\beta|^2=\sin^2\frac{\theta_1}{2}\cos^2\frac{\theta_2}{2}, ~~~  
r=|\gamma|^2=\sin^2\frac{\theta_1}{2}\sin^2\frac{\theta_2}{2}\,,
\eeq
and the $\VEC X_{3,\psi_3}$ is associated to the projector $\VEC P_{\{000,110\}}$. The appearance of the $\VEC X_{3,\psi_3}$ may be surprising but is understandable. Because during the second rotation, not only the components in the projector $\VEC P_{\{011,110\}}$ are generated but also the ones in the projector $\VEC P_{\{000,110\}}$. 

The three azimuthal angles of the $\VEC X$-terms are related to the phase differences of the coefficients: $\psi_1=\pm\arg(\frac{\beta}{\alpha})$, $\psi_2=\pm\arg(\frac{\gamma}{\beta})$ and $\psi_3=\pm\arg(\frac{\gamma}{\alpha})$. The signs are depending on how we choose $\VEC Z$. To fix the sign, we can pick a particular scheme for the ordering of the product states. For example, we can fix the ordering like $\{000\}=\{(1)\}$, $\{011\}=\{(2)\}$ and $\{110\}=\{(3)\}$ and define $\VEC Z_{\{(i),(j)\}} \equiv \{(i)\}-\{(j)\}$ with $i<j$. Then all of the signs are positive. So we have:
\beq
\psi_1 ~=~ \phi_{\beta} - \phi_{\alpha}\,,~~\quad~~  \psi_2 ~=~ \phi_{\gamma} - \phi_{\beta}\,, ~~\quad~~ \psi_3 ~=~  \phi_{\gamma} - \phi_{\alpha} = \psi_1 + \psi_2\,,
\eeq
where $\phi_{\alpha}=\arg(\alpha)$, $\phi_{\beta}=\arg(\beta)$ and $\phi_{\gamma}=\arg(\gamma)$. Therefore, not all of the azimuthal angles of the $\VEC X$'s part are independent just like the phase differences of the $\alpha$, $\beta$ and $\gamma$.

\subsection{General pure states of multi-qubit systems}

The form of the density operator in \eqref{3q_sf} can be generalized to any pure state of multi-qubit systems. However, there are some kinds of structure hidden in the $\VEC X$-terms. Specifically, notice that each coefficient of the $\VEC X_\psi$-terms is always determined by the probabilities of the two states in its corresponding projector. Furthermore, the angles of the $\VEC X$-terms are not independent and constrained somehow. These can all be traced back to the pure state condition: $\rho^2=\rho$. To show this, consider a general density operator written in the following form:
\beq
\label{gs_rho}
\rho ~=~ \sum_{i}\, p_i\, \{(i)\} ~+~ \sum_{i< j}\,q_{i,j}\VEC  X_{\psi_{i,j}}\,,~\quad~ i,j = 1\cdots N\,,
\eeq
where $\{(i)\}$ are the density operators of a series of the orthogonal product states from $\{(1)\}$ to $\{(N)\}$. All of the coefficients in the above expression should be real and nonnegative and $p_i$ should sum to one. We will assume that the definition of $\VEC Z$ is based on the labeled ordering (i.e, $\VEC Z_{\{(i),(j)\}}=\{(i)\}-\{(j)\}$ with $i<j$).  The $\VEC  X_{\psi_{i,j}}$ is associated to the projector $\VEC P_{\{(i),(j)\}}$ and the subscript $\psi_{i,j}$ denote its azimuthal angle. 

%
%\begin{align}
%\label{proj_rules}
%\{(i)\}^2=\{(i)\}\,, ~\quad~ \{(i)\}\{(j)\}=0\,,~\textrm{if}~ i \neq j\,,~&\quad~ \VEC  X_{\psi_{i,j}}^2 = \VEC P_{\{(i),(j)\}} = \{(i)\}+\{(j)\}\,, \nonumber \\
%\{(i)\}\VEC  X_{\psi_{j,k}}  = \VEC  X_{\psi_{j,k}}\{(i)\} = 0\,,~ \textrm{if} ~ i \neq j,k\,, ~&\quad~
%\VEC X_{\psi_{i,j}}\,\VEC  X_{\psi_{k,l}} = 0 ~~\textrm{if} ~ i\,,j \neq k\,,l \,.
%\end{align}

The computation of the $\rho^2$ is greatly simplified by using the projector properties of $\{(i)\}$ and $\VEC P_{\{(i),(j)\}}$ and the \emph{unitarity} of the $\VEC  X$'s. Additionally, because $\VEC  X_{\psi_{i,j}}$ anti-commute with $\VEC Z_{\{(i),(j)\}}$ while commute with $\VEC P_{\{(i),(j)\}}$, the symmetric product between $\{(i)\}$ or $\{(j)\}$ with $\VEC X_{\psi_{i,j}}$ satisfy the following rules:
\beq
\label{Xtimesi}
\{(i)\}\,\VEC  X_{\psi_{i,j}} + \VEC  X_{\psi_{i,j}}\{(i)\} ~=~\{(j)\}\,\VEC  X_{\psi_{i,j}} + \VEC X_{\psi_{i,j}}\{(j)\} ~=~ \VEC X_{\psi_{i,j}}\,\VEC P_{\{(i),(j)\}}= \VEC X_{\psi_{i,j}}\,.
\eeq
Lastly, the only nontrivial product between two different $\VEC  X$'s is when their projector spaces are overlapping. In that case, we can show that the symmetric product of two $\VEC X$'s satisfy the following rule:
\beq
\label{XtimesX}
\VEC X_{\xi_{i,j}}\,\VEC X_{\xi_{j,k}}+\VEC X_{\xi_{j,k}}\,\VEC X_{\xi_{i,j}}=\VEC X_{\xi_{i,k}}\,,
\eeq
where $\VEC X_{\xi_{i,k}}$ is associated to $\VEC P_{\{(i),(k)\}}$ and the angle $\xi_{i,k}$ is:
\beq
\xi_{i,k} ~=~ \sign(j-i)\,\xi_{i,j} + \sign(k-j)\,\xi_{j,k}\,. 
\eeq
To show this, we will use an example. Consider $\{(i)\}=\{00000\}$, $\{(j)\}=\{10011\}$ and $\{(k)\}=\{11100\}$. If $i<j<k$, we have:
\begin{align}
\VEC X_{\xi_{i,j}}\,\VEC X_{\xi_{j,k}}+\VEC X_{\xi_{j,k}}\,\VEC X_{\xi_{i,j}} &= \hx_{a,\xi_{i,j}}\hxd\hxe\,\{(j)\}\,\hx_{b,\xi_{j,k}}\hxc\hxd\hxe+\hx_{b,\xi_{j,k}}\hxc\hxd\hxe\,\{(j)\}\,\hx_{a,\xi_{i,j}}\hxd\hxe \nonumber \\
&= \hx_{a,\xi_{i,j}}\hx_{b,\xi_{j,k}}\hxc\,(\{(i)\}+\{(k)\}) = \VEC X_{\xi_{i,k}}\,,
\end{align}
and $\xi_{i,k} = \xi_{i,j}+\xi_{j,k}$. One can always show that \eqref{XtimesX} is true by adjusting the overlapping $\hx$'s to be in the same direction to cancel them and find the angle of the new $\VEC X$.

Before we compute the square of the $\rho$, notice that the number of the angles in \eqref{gs_rho} is much larger than the degrees of freedom that a pure state should have. This implies there should be some constraints on those parameters if the density operator is a pure state. We will assume the following constraints:
\beq
\label{angle_ps}
\psi_{i,k} ~=~ \sign(j-i)\,\psi_{i,j} + \sign(k-j)\,\psi_{j,k}\,, ~\quad~ i\neq j \neq k\,.
\eeq 
This implies:
\beq
\label{Xeq_ps}
\VEC X_{\psi_{i,k}} \VEC X_{\psi_{j,k}} +\VEC X_{\psi_{j,k}} \VEC X_{\psi_{i,k}} ~=~ \VEC X_{\psi_{i,j}} \,, ~~\quad~~\textrm{for}~~\textrm{any}~~k\neq i,j\,.
\eeq

Using the above assumption about angles and combining the properties of the projectors and the $\VEC X$'s, it is straightforward to compute $\rho^2$ and compare the result to $\rho$. The pure state condition (i.e., $\rho^2=\rho$) require:
\bea
p_i &=& p_i^2 + \sum_{j\neq i}\,q_{i,j}^2\,, \\
q_{i,j} &=& (p_i + p_j)\,q_{i,j}  ~+~ \sum_{k\neq i, j}\,q_{i,k}\,q_{k,j}\,.
\eea
By using the fact that $p_i$ sum to one, we can show that these equalities are satisfied if $q_{i,j}$ equal to $\sqrt{p_i p_j}$. The meaning of the assumption in \eqref{angle_ps} is to ensure that the angles $\psi_{i,j}$ can be parametrized to $\psi_j-\psi_i$ or $\psi_i-\psi_j$. Then, the $\psi_i$ can be understood as the phase of the coefficient of the state $|i\ra$. The total sum of the angles $\psi_i$ are related to the global phase and it will not affect the density operator.

From the above discussion, we can summary that if the following conditions are satisfied
\beq
\label{ps_cond}
q_{i,j}=\sqrt{p_i p_j}\,, ~~\quad~~ \VEC X_{\psi_{i,k}} \VEC X_{\psi_{j,k}} +\VEC X_{\psi_{j,k}} \VEC X_{\psi_{i,k}} ~=~ \VEC X_{\psi_{i,j}} \,,~\quad~ i\neq j \neq k
\eeq
the density operator in \eqref{gs_rho} represent the pure state: $\sum_i\,\alpha_i\, |i\ra$ with:
\beq
p_i = |\alpha_i|^2\,, ~~\quad~~ \psi_{i,j} = \left( \arg(\alpha_j)-\arg(\alpha_i) \right) ~~\textrm{for}~~i<j\,.
\eeq
For a mixed state, the above conditions \eqref{ps_cond} no longer hold and it can be basically regarded as a state with some of the $\VEC X$'s terms missing or weakening.

%%%%%%%%%%%%%%%%%%%%%%%%%%%%%%%%%%%%%%%%%%
\section{Bell states and Bell inequalities}

The Bell states are four specific maximally entangled states of two qubits:
\bea
|\Phi^+\ra = \frac{1}{\sqrt{2}}\,\left(|0\ra_a \otimes |0\ra_b + |1\ra_a \otimes |1\ra_b \right)\,, &\quad |\Phi^-\ra = \frac{1}{\sqrt{2}}\,\left(|0\ra_a \otimes |0\ra_b - |1\ra_a \otimes |1\ra_b \right)\,, \\
|\Psi^+\ra = \frac{1}{\sqrt{2}}\,\left(|0\ra_a \otimes |1\ra_b + |1\ra_a \otimes |0\ra_b \right)\,, &\quad |\Psi^-\ra = \frac{1}{\sqrt{2}}\,\left(|0\ra_a \otimes |1\ra_b - |1\ra_a \otimes |0\ra_b \right)\,,
\eea
where $|0\ra_a \otimes |0\ra_b$ is corresponding to the product state $\{00\}$ and etc. Since they are maximally entangled (i.e., $\theta=\pi/2$), they should be located at the equator of the (entanglement) sphere. From this and the projector space they should be in, it is straightforward to find out their density operators:
\begin{align}
|\Phi^+\ra ~~&\rightarrow~~ \rho_{\Phi^+} =\frac{1}{2}\,(\VEC P_{\{00,11\}} + \VEC X_{\{00,11\}}) = \frac{1}{4}\,(1 + \hza\hzb + \hxa\hxb - \hya\hyb)\,, \\
|\Phi^-\ra ~~&\rightarrow~~ \rho_{\Phi^-} =\frac{1}{2}\,(\VEC P_{\{00,11\}} - \VEC X_{\{00,11\}}) = \frac{1}{4}\,(1 + \hza\hzb - \hxa\hxb + \hya\hyb)\,, \\
|\Psi^+\ra ~~&\rightarrow~~ \rho_{\Psi^+} =\frac{1}{2}\,(\VEC P_{\{01,10\}} + \VEC X_{\{01,10\}}) = \frac{1}{4}\,(1 - \hza\hzb + \hxa\hxb + \hya\hyb)\,, \\
|\Psi^-\ra ~~&\rightarrow~~ \rho_{\Psi^-}  =\frac{1}{2}\,(\VEC P_{\{01,10\}} - \VEC X_{\{01,10\}}) = \frac{1}{4}\,(1 - \hza\hzb - \hxa\hxb - \hya\hyb)\,.
\end{align}

Note that all of the Bell states are related by some local rotations. For example, if we rotate the qubit $a$ around the $y$-axis by $\pi$, then $\hx_a\rightarrow-\hx_a$ and $\hz_a\rightarrow-\hz_a$. Therefore, $|\Psi^-\ra \rightarrow |\Phi^+\ra$. On the other hand, $|\Psi^-\ra$ is a singlet state and it is rotational invariant. This can be seen from the fact that the term: $\hxa\hxb+\hya\hyb+\hza\hzb$ is invariant under the joint action of the local rotors:
\bea
R_{\uVEC n_a, \hza}\,R_{\uVEC n_b, \hzb}\,(\hxa\hxb+\hya\hyb+\hza\hzb) R^{\dag}_{\uVEC n_b, \hzb}\,R^{\dag}_{\uVEC n_a, \hza}\, &=& \uVEC n_{a\dashv}  \uVEC n_{b\dashv} + \uVEC n_{a\vdash}  \uVEC n_{b\vdash} + \uVEC n_a \uVEC n_b \\
&=& \hxa\hxb+\hya\hyb+\hza\hzb\,,
\eea
where $R_{\uVEC n_a, \hza}$ and $R_{\uVEC n_b, \hzb}$ are copies of the same rotor but acting on the spaces of the qubit $a$ and $b$ respectively and $(\uVEC n_{\dashv}, \uVEC n_{\vdash}, \uVEC n)$ form the new basis after the space rotation. 

An interesting feature of the singlet state is that the measurements of the two qubits are always anti-correlated in any direction. To see this, we can consider the projective measurement along some direction $\uVEC s$ on the qubit `$a$' and check how the reduced density operator of the qubit `$b$' change. Before the measurement, it is completely mixed (i.e., $\rho_b = \half$). If the result is positive, the post measurement state is
\beq
\rho^\prime_{ab} ~=~ 2\,\left(\frac{1 + \uVEC s_a}{2}\,\frac{1 - \hza\hzb - \hxa\hxb - \hya\hyb}{4}\,\frac{1 + \uVEC s_a}{2}\right)\,,
\eeq
the factor of $2$ coming from dividing the probability $\half$ which is needed for restoring the unit trace. The reduced density operator of the qubit $b$ after the measurement is:
\beq
\rho^\prime_b ~=~ 2\,\la\rho^\prime_{ab}\ra_a ~=~ 2\left\la (1 + \uVEC s_a)\left(\frac{1 - \hza\hzb - \hxa\hxb - \hya\hyb}{4}\right) \right\ra_a ~=~ \frac{1 - \uVEC s_b}{2}\,.
\eeq
Therefore, originally a completely mixed state but after the measurement,  the reduced density operator of the qubit $b$ become a pure state with its spin point at the opposite direction.

The correlations of the measurements on the maximally entangled states are very different from the usual correlations in a classical system. This was shown by the violation of the Bell inequalities which are supposed to be satisfied by any physical theory of local hidden variables (e.g., classical physics). There are several variations of Bell inequalities. The most well known is the CHSH inequality. It can be explained by an experiment performed on two systems, `$a$' and `$b$', with maybe some correlations between them due to their shared history. In the experiment, a series of measurements on each of them is performed by choosing randomly from two different observables, e.g., Q and R for the system `$a$' and S and T for `$b$'. Each measurement has two possible outcomes, either $-1$ or $1$. If the two systems are separated far apart so that their measurements cannot affect each other and assume local realism, it was shown that the following inequality must be satisfied: 
\beq
E(Q\,S) + E(R\,S) + E(R\,T) - E(Q\,T) \leq 2\,,
\eeq
where $E(\cdot)$ denote the mean value of that quantity. 

However, it was shown that this inequality can be violated by some quantum states. To check that in the MSTA, consider the two qubits `$a$' and `$b$' in the singlet state and we measure their spins in the directions of $\uVEC q$ and $\uVEC r$ for `$a$' and $\uVEC s$ and $\uVEC t$ for `$b$'. The expectation value of the $\uVEC q\, \uVEC s$ is:
\beq
4\,\la\,\uVEC q_a\, \uVEC s_b\, \rho_{\Psi^-}\,\ra = 4\,\left\la \uVEC q_a\, \uVEC s_b\, \frac{1 - \hza\hzb - \hxa\hxb - \hya\hyb}{4}\right\ra = -\uVEC q \cdot \uVEC s\,.
\eeq
Therefore, the left hand side of the inequality is:
\beq
4\,\la\, (\uVEC q\, \uVEC s + \uVEC r\, \uVEC s + \uVEC r\, \uVEC t - \uVEC q\, \uVEC t)\,\rho_{\Psi^-}\,\ra = - ( \uVEC r + \uVEC q) \cdot \uVEC s - ( \uVEC r - \uVEC q) \cdot \uVEC t\,.
\eeq
To maximize this value, we can simply choose $\uVEC s$ and $\uVEC t$ along the opposite directions of $(\uVEC r + \uVEC q)$ and $(\uVEC r - \uVEC q)$ respectively. The result is:
\beq
|\uVEC r + \uVEC q| ~+~ |\uVEC r - \uVEC q| = \sqrt{2+2\,\cos\theta} ~+~\sqrt{2 - 2\,\cos\theta}\,,
\eeq
where $\theta$ is the angle between $\uVEC r$ and $\uVEC q$. The maximum is $2\sqrt{2}$ when $\theta=\pi/2$. Clearly, the singlet state violate the CHSH inequality. How about the other Bell states? Since the other maximally entangled states are related to the singlet state by some local rotations, the expectation values of the measurement results can be written as:
\beq
\la R^\dag_b\,R^\dag_a\,(\uVEC q_a \uVEC s_b + \uVEC r_a \uVEC s_b + \uVEC r_a \uVEC t_b - \uVEC q_a \uVEC t_b)\,R_a\,R_b\, \rho_{\Psi^-}\, \ra =  - ( \uVEC r^\prime + \uVEC q^\prime) \cdot \uVEC s^\prime - ( \uVEC r^\prime - \uVEC q^\prime) \cdot \uVEC t^\prime\,,
\eeq 
where $R_{a,b}$ are the rotors of the local rotations and we have used the the cyclic nature of the operation $\la\_\ra$. Clearly the maximum of this term is still $2\sqrt{2}$ but the set of the measurement that will achieve this value will depend on the state. How about the partially entangled states? Since only the terms that contain vectors from both qubits will contribute to the inequality and those terms are smaller for the partially entangled state, the state will violate the inequality less strongly when the entanglement angle $\theta$ become smaller. And, the inequality will be satisfied for any product state. How much the Bell inequality can be violated by a quantum state is limited by the Tsirelson bound. For the CHSH inequality, this bound is $2\sqrt{2}$ and we have shown that this is true for a two-qubit system.

%%%%%%%%%%%%%%%%%%%%%%%%%%%%%%%%%%%%%%%%%%%%%%%%%%
\section{Dynamics of two coupled qubits}

%Hamiltonian

%In this section, we consider how the density operator evolve with time when the two qubits are interacting with each other. 

To show how the MSTA formulation can help us analyze quantum systems in a geometrical way, we use the dynamics of two coupled qubits as our examples. The time evolution of a density operator is determined by the Hamiltonian $\VEC H$ of the system:
\beq
\rho(t)=e^{-\iota \VEC H t}\,\rho(0)\,e^{\iota \VEC H t}\,.
\eeq
Notice that this is similar to how we applying a rotor on the density operator. Indeed, the time evolution of a pure state (for a closed system) is basically a unitary transformation and as we have explained, can be regarded as the action of a rotor or a series of the rotors. This provides some geometrical perspectives on the dynamics of quantum systems. We will show some simple cases with the two-qubit systems in which the Hamiltonian can be realized as a single rotor or two separated rotors. 

\subsection{Isotropic interaction}

We consider the isotropic exchange interaction between two qubits. The Hamiltonian in the MSTA is:
\beq
\label{isoH}
\VEC H = \frac{\omega}{4}\,(\hxa\hxb+\hya\hyb+\hza\hzb)\,,
\eeq
where $\omega$ determine the strength of the coupling. We can rewrite the Hamiltonian in the following form:
\bea
\VEC H &=& \frac{\omega}{4}\,(\hxa\hxb+\hya\hyb+\hza\hzb)\,\left(\VEC P_{\{00,11\}} +\VEC P_{\{01,10\}}\right)\,, \\
&=& \frac{\omega}{4}\,(\VEC P_{\{00,11\}} - \VEC P_{\{01,10\}} + 2\,\VEC X_{\{01,10\}})\,.
\eea

The projector terms in the Hamiltonian will do nothing to the states in the projector $\VEC P_{\{00,11\}}$ or $\VEC P_{\{01,10\}}$. The time evolution operator is clearly a rotor in the projector $\VEC P_{\{01,10\}}$ around the axis $\VEC X_{\{01,10\}}$. For example, if the initial density operator $\rho(0)$ is any state in $\VEC P_{\{01,10\}}$, its time evolution will be:
\beq
e^{-\iota\, \VEC X_{\{01,10\}}\,(\omega t/2)}\,\rho_{\VEC S}(0)\,e^{\iota\, \VEC X_{\{01,10\}}\,(\omega t/2)} = \frac{1}{2}\,(\VEC P_{\{01,10\}} + e^{-\iota\, \VEC X_{\{01,10\}}\,(\omega t/2)} \VEC S_{\{01,10\}}\,e^{\iota\, \VEC X_{\{01,10\}}\,(\omega t/2)})\,.
\eeq
Therefore, the dynamic can be pictured as the rotation of a unit vector around the axis $\VEC X_{\{01,10\}}$. On the other hand, if the initial density operator is in $\VEC P_{\{00,11\}}$, then it will not change with time. From these observations, we can see that the Hamiltonian preserve the structure of the projectors and the reason is that it commutes with the projectors:
\beq
\VEC H\,\VEC P_{\{00,11\}} =\VEC P_{\{00,11\}}\, \VEC H\,, ~\quad~ \VEC H\,\VEC P_{\{01,10\}} =\VEC P_{\{01,10\}}\, \VEC H\,.
\eeq
In other words, the time evolution operator can be regarded as a rotor in some projective space only when the Hamiltonian commutes with the particular projector. Then, we can use that projector to project the Hamiltonian to obtain the corresponding rotor. 

Additionally, from the above analysis, we can conclude the two of the eigenstates of this Hamiltonian in the projector $\VEC P_{\{01,10\}}$ are those states along the rotation axis (i.e., the two Bell states, $\rho_{\Psi^+}$ and $\rho_{\Psi^-}$) and the other two are any two orthogonal states in $\VEC P_{\{00,11\}}$.  

Since the Hamiltonian is isotropic, the time evolution of any state in which the spins of the two qubits are aligned or anti-aligned can be pictured in this way. If the spins are neither aligned nor anti-aligned initially, their directions will change with time. We will talk about this more complicated case later.

\subsection{Anisotropic interaction}

We can also consider the anisotropic exchange interaction between two qubits. The Hamiltonian in the MSTA can be written as:
\beq
\VEC H = \frac{1}{4}\,(\omega_x\,\hxa\hxb + \omega_y\,\hya\hyb + \omega_z\, \hza\hzb)\,,
\eeq
where $\omega_x$, $\omega_y$ and $\omega_z$ control the coupling in different directions. This Hamiltonian can also be projected by the same projectors:
\bea
\VEC H &=& \frac{1}{4}\,(\omega_x\,\hxa\hxb + \omega_y\,\hya\hyb + \omega_z\, \hza\hzb)\,\left(\VEC P_{\{00,11\}} +\VEC P_{\{01,10\}}\right)\,  \nonumber \\
&=& \frac{1}{4}\,( (\omega_x-\omega_y)\,\VEC X_{\{00,11\}} + \omega_z\,\VEC P_{\{00,11\}})
+ \frac{1}{4}\,( (\omega_x+\omega_y)\,\VEC X_{\{01,10\}} - \omega_z\,\VEC P_{\{01,10\}})\,.
\eea
Notice that the two parts commute and the time evolution will be regarded as the separated rotations in the two projector spaces. For example, for the initial state $\{00\}$ or $\{01\}$, the time evolution will be:
\bea
\rho(t) &=& e^{-\iota\, \VEC X_{\{00,11\}}\,(\omega_- t/2)}\,\{00\}\,e^{\iota\, \VEC X_{\{00,11\}}\,(\omega_- t/2)} \, \nonumber \\
&=& \frac{1}{2}\,( \VEC P_{\{00,11\}} + \cos(\omega_-t)\,\VEC Z_{\{00,11\}} - \sin(\omega_-t)\,\VEC Y_{\{00,11\}})\,, \\
\rho(t) &=& e^{-\iota\, \VEC X_{\{01,10\}}\,(\omega_+ t/2)}\,\{01\}\,e^{\iota\, \VEC X_{\{01,10\}}\,(\omega_+ t/2)}\, \nonumber \\
&=& \frac{1}{2}\,( \VEC P_{\{01,10\}} + \cos(\omega_+t)\,\VEC Z_{\{01,10\}} - \sin(\omega_+t)\,\VEC Y_{\{01,10\}})\,,
\eea
where $\omega_{\pm} \equiv (\omega_x \pm \omega_y)/2$. From the above evolution, we can also see how the entanglement develop under interaction. Since the entanglement angle $\theta$ equal to $\omega_{\pm}t$, the entanglement between two qubits change periodically. The period is depending on the initial state and it is $\pi/\omega_-$ for $\{00\}$ and $\pi/\omega_+$ for $\{01\}$.

The time evolution of any state in which the spins of the two qubits are aligned or anti-aligned in the directions of $\hx$, $\hy$ or $\hz$ can all be pictured in this way. The reason is that the Hamiltonian commutes with the corresponding projector. For the other initial states, one may need to do the direct computation or consider the method described in section 5.4.

\subsection{Anisotropic interaction with an external field} 

We can also include some external magnetic field along the $\hz$ in the previous example. For example, we can consider the following Hamiltonian:
\beq
\VEC H = \frac{1}{4}\,(\omega_x\,\hxa\hxb + \omega_y\,\hya\hyb + \omega_z\, \hza\hzb) + \frac{1}{2}\,(\beta_a\, \hza + \beta_b\, \hzb)\,,
\eeq
where $\beta_a$ and $\beta_b$ are related to the strength of the magnetic field and the couplings of two qubits with the filed.  This Hamiltonian can also be projected by the same projectors:
\beq
\VEC H =  \frac{1}{2}\,\left( \omega_-\,\VEC X_{\{00,11\}} + \beta_+\,\VEC Z_{\{00,11\}} + \frac{\omega_z}{2}\,\VEC P_{\{00,11\}} \right)
+ \frac{1}{2}\,\left(  \omega_+\, \VEC X_{\{01,10\}} +\beta_-\,\VEC Z_{\{01,10\}} - \frac{\omega_z}{2}\,\VEC P_{\{01,10\}} \right)\,,
\eeq
where $\beta_\pm \equiv \beta_a \pm \beta_b$. Again, the time evolution can be regarded as the separated rotations in the two projector spaces but around the different axes this time. For the states in $\VEC P_{\{00,11\}}$, the axis is along:
\beq
\VEC A_{\{00,11\}} = \frac{\omega_-}{\omega_{00}} \VEC X_{\{00,11\}} + \frac{\beta_+}{\omega_{00}} \VEC Z_{\{00,11\}}\,,
\eeq
where $\omega_{00} \equiv \sqrt{\omega_-^2 + \beta_+^2}$ is the angular speed of the rotation. For the states in $\VEC P_{\{01,10\}}$, the axis is along:
\beq
\VEC A_{\{01,10\}} = \frac{\omega_+}{\omega_{01}} \VEC X_{\{01,10\}} + \frac{\beta_-}{\omega_{01}} \VEC Z_{\{01,10\}}\,,
\eeq
where $\omega_{01} \equiv \sqrt{\omega_+^2 + \beta_-^2}$ is the angular speed of this rotation. 

From the rotation axes, we can see that how the entanglement develop will depend on the relative strength between $\beta_+$ and $\omega_-$ or $\beta_-$ and $\omega_+$. If $\beta_+ \gg \omega_-$ or $\beta_- \gg \omega_+$, the rotation axis will be very close to $\VEC Z$ and the entanglement in that corresponding projector will not change much.

Additionally, we can conclude from the above analysis that the four eigenstates of the Hamiltonian are:
\beq
\rho_1^{(\pm)} ~=~ \frac{1}{2}\,(\VEC P_{\{00,11\}} \pm \VEC A_{\{00,11\}})\,, ~\quad~ \rho_2^{(\pm)} ~=~ \frac{1}{2}\,(\VEC P_{\{01,10\}} \pm \VEC A_{\{01,10\}})\,.
\eeq
Their corresponding eigenvalues are:
\beq
4\,\la \VEC H\, \rho_1^{(\pm)} \ra= \frac{\omega_z \pm 2\,\omega_{00}}{4}\,, ~\quad~ 4\,\la \VEC H\, \rho_2^{(\pm)} \ra= \frac{-\omega_z \pm 2\,\omega_{01}}{4}\,.
\eeq

\subsection{Starting from a general initial product state} 

We have seen that the time evolution of the states in the appropriate projectors (i.e., those commute with the Hamiltonian) can be pictured as some simple rotations. For the other states, we may need to do the direct computation to find out the evolution. However, we will show that it is possible (at least in some cases) to break the initial density operator to several pieces which can be identified as some elements in the appropriate projectors. Then, the evolution of the state can be obtained as the result of the combined rotations of all pieces.

The example we are going to use is the isotropic interaction in \eqref{isoH} with the following initial state:
\beq
\rho(0) = \left(\frac{1+\uVEC m_a}{2}\right)\left(\frac{1+\uVEC n_b}{2}\right)\,,
\eeq
where $\uVEC m$ and $\uVEC n$ are two arbitrary unit vectors. We can rewrite the initial density operator in this form:
\beq
\rho(0) = \frac{1}{4}\,(1+ (\vt p_a + \vt q_a) + (\vt p_b - \vt q_b) + (\vt p_a\, \vt  p_b - \vt q_a\, \vt q_b + \vt q_a\, \vt p_b - \vt p_a\, \vt q_b))\,,
\eeq
where $\vt p_{a,b}$ and $\vt q_{a,b}$ are:
\beq
\vt p_{a,b} \equiv \frac{\uVEC m_{a,b} + \uVEC n_{a,b}}{2} = p\,\uVEC p_{a,b}\,, ~\quad~ \vt q_{a,b} \equiv \frac{\uVEC m_{a,b} - \uVEC n_{a,b}}{2} = q\,\uVEC q_{a,b}\\,.
\eeq
Note that $\uVEC m_b$ point at the same direction as $\uVEC m_a$ but it is a vector in the qubit $b$ space and similarly for $\uVEC n_a$. 

Ignoring the scalar, we can break the initial density operator to three parts and deal with them in three different projectors:
\beq
(\VEC A) = \frac{1}{4}\,(\vt p_a + \vt p_b + \vt p_a\,\vt p_b)\,, ~\quad~ (\VEC B) = \frac{1}{4}\,(\vt q_a - \vt q_b - \vt q_a\,\vt q_b)\,, ~\quad~ (\VEC C) = \frac{1}{4}\,(\vt q_a\, \vt p_b - \vt p_a\,\vt q_b)\,,
\eeq
Due to their resemblance with the aligned and anti-aligned product states, the evolution of the first two parts can be easily seen by using the projectors: $(1\pm\uVEC p_a\,\uVEC p_b)/2$ and $(1\pm\uVEC q_a\,\uVEC q_b)/2$ to expand the Hamiltonian separately. The part ($\VEC C$) can be regarded as a component in the following projector:
\beq
(\VEC C) = \frac{p\,q}{4}\,\uVEC q_a\,\uVEC p_b\, \VEC P_{\{+-,-+\}}\,, ~\quad~ \VEC P_{\{+-,-+\}} = \frac{1-\uVEC r_a\,\uVEC r_b}{2}\,.
\eeq
where $\uVEC r_{a,b} \equiv -\iota\,\uVEC p_{a,b} \uVEC q_{a,b}$. 

Using what we have learned before, we find out that for the isotropic interaction, only the following parts changed with time:
\bea
q\,\frac{\uVEC q_a - \uVEC q_b}{2} ~&\rightarrow&~ q\,\left(\cos(\omega t)\left(\frac{\uVEC q_a - \uVEC q_b}{2}\right) - \sin(\omega t)\left(\frac{\uVEC r_a \uVEC p_b - \uVEC p_a \uVEC r_b}{2}\right) \right)\,, \\
p\,q\,\frac{\uVEC q_a \uVEC p_b - \uVEC p_a \uVEC q_b}{2} ~&\rightarrow&~ p\,q\,\left( \cos(\omega t)\left(\frac{\uVEC q_a \uVEC p_b - \uVEC p_a \uVEC q_b}{2} \right) + \sin(\omega t)\left(\frac{\uVEC r_a - \uVEC r_b}{2}\right) \right)\,.
\eea
The time evolution for the case with the anisotropic interaction or the other cases introduced before can also be found out in a similar way.

From the above result, we can find out the reduced density operators for the two qubits:
\bea
\rho_a = \frac{1}{2}\,(1 + \vt p_a + \cos(\omega t)\, \vt q_a + p\,q\,\sin(\omega t)\,\uVEC r_a )\,, \\
\rho_b = \frac{1}{2}\,(1 + \vt p_b - \cos(\omega t)\, \vt q_b - p\,q\,\sin(\omega t)\,\uVEC r_b )\,. 
\eea
We can see clearly that the two spins are circling around the axis along $\uVEC p$. Furthermore, notice that the lengths of the vectors of both qubits are not preserved during the process. The means the two qubits entangle a bit with each other periodically. The maximum entanglement they can achieve (which happen at $\omega t=\pi/2,3\pi/2,\cdots$) depends on the angle between the two spin vectors \emph{initially}. We can compute the shortest length of the vector of one of the qubits during the process and the result is:
\beq
\frac{1}{4}\,\sqrt{5+8\cos\psi+2\cos\psi^2+\cos\psi^4}\,,  ~\quad~ \uVEC m \cdot \uVEC n =\cos\psi\,.
\eeq
This function monotonically decreases from one to zero when we change the angle $\psi$ from zero to $\pi$. This means that the two qubits get entangled more if the angle between their spins is larger and when the directions are anti-aligned, the entanglement can reach the maximum.

%%%%%%%%%%%%%%%%%%%%%%%%%%%%%%%%%%%%%%%%%%%%%%%%%%%
\section{The local invariants and the space of two-qubit and three-qubit pure states}

One of the advantages of using the MSTA formulation is the transparency of the effects of the local unitary transformations. From the geometric perspective, the local unitary transformation of a particular qubit can be regarded as some rotation of the Bloch sphere of that qubit. Then, because the vectors or bi-vectors of the different qubits will commute with each other, the local operations simply rotate all of the involving vectors in the density operator separately. This provides a natural way to define the local unitary invariants by extracting the scalar part of the products of the several components in the density operator. Then, these invariants can be used to parametrize pure states in a local-unitary invariant way.

To show how this work practically, we will use two-qubit and three-qubit pure states as our examples. We will define the canonical coordinates\footnote{A similar idea for three-qubit pure states in matrix formalism was described in \cite{sudbery2001local}. On the other hand, a different method based on the Schmidt decomposition to define the canonical forms was introduced in \cite{acin2001three}.} for the pure states by using the invariants introduced naturally in the MSTA formulation. The coordinates provide some sort of the local invariant labels for these states. This enables us to study the space of these states with the local and non-local degrees of freedom separated.

\subsection{Two-qubit pure states}

The density operator of any state of the two-qubits $a$ and $b$ can be written in this form:
\beq
\label{rho_ab}
\rho_{ab} ~=~ \frac{1}{4}\,( 1+{\vt v_a}+{\vt v_b}+{\VEC V_{ab}})\,,
\eeq
where ${\VEC V_{ab}}$ is the sum of several terms containing the vectors of both qubits. Then, the scalar part of any product of several components will be a local unitary invariant. For example, ${\vt v_a^2}$, ${\vt v_b^2}$, $\la {\vt v_a}{\vt v_b}{\VEC V_{ab}} \ra$, $\la {\VEC V_{ab}^2} \ra$ and etc. To use these invariants to define the canonical coordinates for a pure state, we can expand the state by a particular set of the product states tailored to it. For example, for the state in \eqref{rho_ab}, we will use the following four product states:
\beq
\{i\,j\} ~=~  \left(\frac{1 + i\, \uVEC v_a}{2}\right) \left( \frac{1 + j\, \uVEC v_b}{2} \right)\,, ~\quad~ i,j=\pm.
\eeq
where $\uVEC v_a = \vt v_a/v_a$ and $\uVEC v_b = \vt v_b/v_b$. The values $v_a$ and $v_b$ are the lengths of the corresponding vectors. For those states without the grade-one components (i.e., $\vt v_a$ and $\vt v_b$), we will discuss them separately.

If the state is pure, then the density operator can be put in the following form:
\beq
\label{2q_gf}
\rho_{ab} ~=~ \sum_{i,j=+,-}\,p_{\{ij\}}\,\{i\,j\} ~+~ \sum_{ij \neq kl} \sqrt{p_{\{ij\}}p_{\{kl\}}}\,\VEC X_{\{ij\,,kl\}}\,,
\eeq
where the four expansion probabilities are:
\beq
p_{\{ij\}} ~=~ 4\,\la\,\{i\,j\}\,\rho_{ab}\,\ra ~=~ \frac{1}{4}\,(1+ i\,v_a + j\,v_b + i\,j\,\la \uVEC v_a \uVEC v_b \VEC V_{ab} \ra )\,, ~\quad~ i,j=\pm.
\eeq
If both of the values $v_a$ and $v_b$ are not zero, the last term can be expressed as: 
\beq
\la \uVEC v_a \uVEC v_b \VEC V_{ab} \ra = \frac{\la \vt v_a \vt v_b \VEC V_{ab} \ra}{v_a\,v_b}\,.
\eeq
Therefore, the expansion probabilities are parametrize by the three invariants $v_a$, $v_b$ and $\la \vt v_a \vt v_b \VEC V_{ab} \ra$. However, these invariants are not independent and related to each other (if the state is pure). Their relations can be verified by checking the following consistent condition. By expanding the first part (i.e., diagonal components) in \eqref{2q_gf}, we notice that both of the $\vt v_a$ and $\vt v_b$ in the density operator are coming from it. The consistency require that either the second part (i.e., the $\VEC X$'s part) produce no grade-one vector  of any qubit or all of that produced are cancelled with each other. However, in general there are four terms in \eqref{2q_gf} that can produce some grade-one vectors:
\bea
\VEC X_{\{++,-+\}} = \uVEC v_{a\perp}\, \left( \frac{1 + \uVEC v_b}{2} \right)\,, ~~\quad~~ \VEC X_{\{+-,--\}} = \uVEC v'_{a\perp}\, \left( \frac{1 - \uVEC v_b}{2} \right)\,, \\
\VEC X_{\{++,+-\}} = \uVEC v_{b\perp}\, \left( \frac{1 + \uVEC v_a}{2} \right)\,, ~~\quad~~ \VEC X_{\{-+,--\}} = \uVEC v'_{b\perp}\, \left( \frac{1 - \uVEC v_a}{2} \right)\,,
\eea
where $\uVEC v_{a\perp}\,,\uVEC v'_{a\perp}\,, \uVEC v_{b\perp}$ and $\uVEC v'_{b\perp}$ are some unit vectors orthogonal to the $\uVEC v_a$ or $\uVEC v_b$ respectively. These vectors need to be cancelled in order to be consistent. Therefore, the following conditions are required:
\bea
\uVEC v'_{a,\perp} = -\uVEC v_{a,\perp}\,, ~~&\quad&~~ \uVEC v'_{b,\perp} = -\uVEC v_{b,\perp}\,, \\
p_{\{++\}}\,p_{\{-+\}} = p_{\{+-\}}\,p_{\{--\}} \,,~~&\quad&~~ p_{\{++\}}\,p_{\{+-\}} = p_{\{-+\}}\,p_{\{--\}}\,.
\eea
From the conditions involves the probabilities, we can see that the non-trivial solutions (i.e., without vanishing $v_a$ and $v_b$) exist only when both $p_{\{+-\}}$ and $p_{\{-+\}}$ vanish. This means:
\beq
v_a ~=~ v_b ~\equiv~ v\,, ~~\quad~~ \la \vt v_a \vt v_b \VEC V_{ab} \ra ~=~ v_a\,v_b ~=~ v^2\,.
\eeq
Therefore we find out that only two of the probabilities, $p_{++}$ and $p_{--}$, remain and they are parametrized by only one invariant $v$. This result agree with what we expect from the Schmidt decomposition. The density operator is reduced to:
\beq
\label{rho12_red}
\rho_{ab} = \left(\frac{1+v}{2}\right)\{++\} ~+~ \left(\frac{1-v}{2}\right)\{--\} ~+~ \left(\frac{\sqrt{1-v^2}}{2}\right)\VEC X_{\{++,--\}}\,.
\eeq
Since the probabilities must be nonnegative, the range of the value $v$ is between 0 and 1. Besides of this invariant, there is still one degree of freedom remaining in the above density operator and it comes from the direction of $\VEC X_{\{++,--\}}$: 
\beq
\VEC X_{\{++,--\}} ~=~ \uVEC v_{a\dashv}\,\uVEC v_{b\dashv}\,\VEC P_{\{++,--\}}\,,
\eeq
where $\uVEC v_{a\dashv}$ and $\uVEC v_{b\dashv}$ are two unit vectors orthogonal to $\uVEC v_a$ and $\uVEC v_b$. The direction of $\VEC X_{\{++,--\}}$ is controlled by the directions of $\uVEC v_{a\dashv}$ and $\uVEC v_{b\dashv}$ which can be changed by some local rotations. This implies that this degree of freedom is local and not really an invariant. Therefore, the value $v$ is actually the only invariant needed in the two-qubit case.

%%%%%%%%%%%%%%%%%%%%%%%%%%%%%%%%%%%%%%%%%%%%%%%%%%%%%
We can count the degrees of freedom of a two-qubit pure state to verify this. The state is described by four complex numbers with eight degrees of freedom. After removing two of them by applying the normalization condition and discarding the global phase, we still have six degrees of freedom. Four of them are coming from choosing the directions of ${\uVEC v_a}$ and ${\uVEC v_b}$ and one of them is related to the direction of the $\VEC X$ (which are related to the local rotations around the spins of two qubits). Therefore, we have the one remaining as the only invariant. 
%%%%%%%%%%%%%%%%%%%%%%%%%%%%%%%%%%%%%%%%%%%%%%%%%%%

Since we have only one invariant, it should be directly related to the entanglement of the two qubits. Indeed, the value $v$ is basically the length of the vector part in the reduced density operator of each qubit, its relation with the entanglement entropy (i.e., von Neumann entropy) has been explained in \eqref{S_tq}. Furthermore, one can show that it is related to the concurrence \cite{hill1997entanglement}, a well-known entanglement measurement. For a two-qubit pure state, the concurrence can be computed by:
\beq
{\mathcal C}(\rho_{ab}) = \sqrt{2(1-\Tr\rho_a^2)}\,,
\eeq
where $\rho_a$ is the reduced density operator of the qubit $a$. Therefore, the concurrence is equal to $\sqrt{1-v^2}$.

Combining all of the parameters (including the local ones),  the state space is basically a half sphere fibered with two $S_2$ coming from the choices of the directions of ${\uVEC v_a}$ and ${\uVEC v_b}$. The pole of the half sphere is the product state $\{++\}$ and that is when $v$ equal to one. When we approach to the equator, the value $v$ decrease to zero and the state become maximally entangled. If we take the limit of $v\to0$, the density operator in \eqref{rho12_red} becomes:
\beq
\label{rho_max}
\rho_{ab} ~=~ \frac{1}{4}\,(1+ \uVEC v_a \uVEC v_b + \uVEC v_{a\dashv}\uVEC v_{b\dashv} - \uVEC v_{a\vdash} \uVEC v_{b\vdash} )
\eeq
where $\uVEC v_{a\vdash}$ and $\uVEC v_{b\vdash}$ are:
\beq
\uVEC v_{a\vdash} = \iota\, \uVEC v_{a\dashv} \uVEC v_a\,, ~~\quad~~ \uVEC v_{b\vdash} = \iota\, \uVEC v_{b\dashv} \uVEC v_b \,.
\eeq
From the above density operator in \eqref{rho_max},  we can see that the state is related to the Bell states by some local rotations. However, it does not mean that the space of maximally entangled states is topologically $S_2\times S_2 \times S_1$. Because we can verify that some points in that space should be identified. For example, the two choices of the spin directions: $(\uVEC v_a, \uVEC v_b)=(\hza,\hzb)$ and $(\hxa,\hxb)$ will yield the same density operators if we choose their azimuthal angles in $S_1$ accordingly. Therefore the space should be some kind of quotient space. The topology of this space is indeed quite interesting however we will not discuss it further here.

\subsection{Three-qubit pure states}

The three-qubit case is much more complicated than two-qubit one because there are more local unitary invariants. However, we can proceed similarly by choosing the appropriate set of product states to expand the density operator and analyzing the consistent conditions.

\subsubsection{The invariants and the consistent conditions}

The density operator of a general three-qubit state can be written in the following form:
\beq
\rho_{abc}=\frac{1}{8}\,(1 + \vt v_a + \vt v_b + \vt v_c + \VEC V_{ab} + \VEC V_{ac} + \VEC V_{bc} + \VEC V_{abc})\,,
\eeq
where $\VEC V_{ab}$, $\VEC V_{ac}$ and $\VEC V_{bc}$ contain several products of the vectors from two of the qubits while $\VEC V_{abc}$ involves all of them. We will choose the following eight product states to expand the density operator:
\beq
\{ijk\} \equiv \left(\frac{1 + i\, \uVEC v_a}{2}\right)\left(\frac{1 + j\, \uVEC v_b}{2}\right)
\left(\frac{1 + k\, \uVEC v_c}{2}\right)\,, ~\quad~ i,j,k = \pm.
\eeq
%\beq
%\rho_{1\pm} ~=~ \frac{1\pm\hv_1}{2}\,, ~\quad~ \rho_{2\pm} ~=~ \frac{1\pm\hv_2}{2}\,, ~\quad~ \rho_{3\pm} ~=~ \frac{1\pm\hv_3}{2}\,,
%\eeq
where $\hv_a=\vt v_a/v_a\,$, $\hv_b=\vt v_b/v_b$ and $\hv_c=\vt v_c/v_c$. The values $v_a$, $v_b$ and $v_c$ are the lengths of the three vectors. We will assume $\rho_{abc}$ is a pure state and therefore it can be expanded into the following form:
\beq
\label{3q_gf}
\rho_{abc} ~=~ \sum_{i,j,k=+,-}\,p_{\{ijk\}}\,\{i\,j\,k\} ~+~ \sum_{ijk \neq lmn} \sqrt{p_{\{ijk\}}p_{\{lmn\}}}\,\VEC X_{\{ijk\,,lmn\}}\,,
\eeq
where the pure state conditions \eqref{ps_cond} should be satisfied. The eight expansion probabilities are:
\bea
\label{p_cc}
p_{\{ijk\}} &=& 8\la\, \{ijk \}\,\rho_{abc}\, \ra\,,  \nonumber \\ 
&=& \frac{1}{8}(1+ i\, v_a + j\, v_b + k\,v_c + i\, j\,\vb_{ab}
+ i\,k\,\vb_{ac} + j\,k\, \vb_{bc} + i\,j\,k\, \vb_{abc}) \,,
\eea
where the scalars, $\vb_{ab}$, $\vb_{bc}$, $\vb_{ac}$ and $\vb_{abc}$ are defined by:
\beq
 \vb_{ab} \equiv \la \uVEC v_a \uVEC v_b \VEC V_{ab} \ra\,, \quad \vb_{ac} \equiv \la \uVEC v_a \uVEC v_c \VEC V_{ac} \ra\,, \quad \vb_{bc} \equiv \la \uVEC v_b \uVEC v_c \VEC V_{bc} \ra\,,
\quad \vb_{abc} \equiv \la \uVEC v_a \uVEC v_a \uVEC v_a \VEC V_{abc} \ra\,.
\eeq 
If all of the values $v_a$, $v_b$ and $v_c$ are not zero, the above four scalars are all invariants. Therefore, the expansion probabilities are parametrized by the seven invariants. However, they are not independent and the relations between them will be uncovered by analyzing the consistent conditions in a similar way. For those states with one or more of the values $v_a$, $v_b$ and $v_c$ vanishing, we will talk about them later.

By checking the expansion form of the density operator, we again find out that the grade-one part (i.e., $\vt v_a$, $\vt v_b$ and $\vt v_c$) are produced completely from the first part (i.e., the diagonal components) in \eqref{3q_gf}. This means that all of the grade one vectors produced from the second part should be cancelled with each other. There are twelve such terms (i.e, four terms for each qubit) and they produced the following vectors:
\beq
\VEC X_{\{+ij, -ij\}} \rightarrow \hv_{\left[\perp i  j\right]}\,, \quad \VEC X_{\{i+j,i-j\}} \rightarrow \hv_{\left[i \perp j\right]}\,, \quad \VEC X_{\{i j +, i j -\}} \rightarrow \hv_{\left[i j \perp\right]}\,, ~\quad~ i,j=\pm. \nonumber
\eeq
Hopefully the notation is not too confusing. The vector $\uVEC v_{[\perp i j]}$ denotes some unit vector of the qubit $a$ orthogonal to the vector $\uVEC v_a$ while the vectors $\uVEC v_{[i\perp j]}$ and $\uVEC v_{[i j \perp]}$ are the orthogonal unit vectors of the qubit $b$ and $c$ respectively. Therefore, we have four vectors on the plane orthogonal to the spin for each qubit.

For a pure state, the directions of these twelve vectors can not be  chosen independently. They are related by the pure state condition in \eqref{ps_cond}. Between any two of the qubits, we have two constraints on how their vectors related with each others. For example, for the qubits $a$ and $b$, we have:
\beq
\VEC X_{\{+++,-++\}}\,\VEC X_{\{-++,--+\}} ~+~ \leftrightarrow ~=~ \VEC X_{\{+++,+-+\}}\,\VEC X_{\{+-+,--+\}}~+~ \leftrightarrow\,,
\eeq
and this implies:
\beq
(\hv_{\left[\perp + +\right]}\hv_{\left[- \perp +\right]}-\hv_{\left[\perp - +\right]}\hv_{\left[+ \perp +\right]})(\{+++\}+\{--+\})=0\,. 
\eeq
Using what we learned about the effects of the local axial rotations in the projector space, we can see that the above equation implies that the angle between $\hv_{[\perp++]}$ and $\hv_{[\perp-+]}$ should be equal to the angle between $\hv_{[+\perp+]}$ and $\hv_{[-\perp+]}$ (including the sign). We can also check the relation from:
\beq
\VEC X_{\{++-,-+-\}}\,\VEC X_{\{-+-,---\}} ~+~ \leftrightarrow ~=~ \VEC X_{\{++-,+--\}}\,\VEC X_{\{+--,---\}}~+~ \leftrightarrow\,.
\eeq
It implies that the angle between $\hv_{[\perp+-]}$ and $\hv_{[\perp--]}$ should be equal to the angle between $\hv_{[+\perp-]}$ and $\hv_{[-\perp-]}$ (including the sign). For the other pairs, we can follow the similar procedure to obtain the relations between the vectors. We summarize them in the following:
\bea
\label{angle_cond}
\measuredangle \left(\hv_{[\perp++]}\,, \hv_{[\perp-+]} \right) =  \measuredangle \left(\hv_{[+\perp+]}\,, \hv_{[-\perp+]}\right)\,, 
\quad \measuredangle \left(\hv_{[\perp+-]}\,, \hv_{[\perp--]} \right) =  \measuredangle \left(\hv_{[+\perp-]}\,, \hv_{[-\perp-]}\right)\,,   \nonumber \\
\measuredangle \left(\hv_{[\perp++]}\,, \hv_{[\perp+-]} \right) =  \measuredangle \left(\hv_{[++\perp]}\,, \hv_{[-+\perp]}\right)\,, 
\quad \measuredangle \left(\hv_{[\perp-+]}\,, \hv_{[\perp--]} \right) =  \measuredangle \left(\hv_{[+-\perp]}\,, \hv_{[--\perp]}\right)\,,   \nonumber\\
\measuredangle \left(\hv_{[+\perp+]}\,, \hv_{[+\perp-]} \right) =  \measuredangle \left(\hv_{[++\perp]}\,, \hv_{[+-\perp]}\right)\,, 
\quad \measuredangle \left(\hv_{[-\perp+]}\,, \hv_{[-\perp-]} \right) =  \measuredangle \left(\hv_{[-+\perp]}\,, \hv_{[--\perp]}\right)\,. 
\eea
Note that once we fix the directions of all twelve vectors (under the above constraints), the directions of the other $\VEC X$'s are completely fixed by the pure state condition. This means that the state is completely determined by the expansion probabilities and the directions of these twelve vectors.

We can introduce the following non-unitary vectors by including the coefficients of the ${\VEC X}$'s terms:
\bea
\vv_{\left[\perp i  j\right]} \equiv\, \sqrt{p_{\{+ i j\}}\,p_{\{- i j\}}}\,  \hv_{\left[\perp i j\right]}\,, ~&\quad&~ \vv_{\left[i \perp j\right]} \equiv \,\sqrt{p_{\{i+j\}} p_{\{i-j\}}}\, \hv_{\left[i \perp j\right]}\,, \nonumber \\
 \vv_{\left[i j \perp \right]} \equiv \,\sqrt{p_{\{i j +\}} p_{\{i j -\}}}\,\hv_{\left[i j \perp\right]}\,, 
 ~&\quad&~ i,j=\pm. \nonumber
\eea
In order to be consistent, the three sets of the vectors must all sum to zero separately. Therefore, we have the following vector sum equations:
\beq
\label{vector_sum}
\sum_{i,j=\pm}\,\vv_{\left[\perp i j\right]}=0\,,~\quad~ \sum_{i,j=\pm}\,\vv_{\left[i \perp j\right]}=0\,, ~\quad~ \sum_{i,j=\pm}\,\vv_{\left[i j \perp\right]}=0\,.
\eeq
The directions of these vectors are constrained by \eqref{angle_cond} so that the three equations are coupled and not easy to solve. However, by utilizing their angle relations, we can extract the equalities containing only the probabilities (related to the lengths of the vectors). At first, from the vector sum equations of the qubit $a$ and $b$, we have:
\bea
(\vv_{\left[\perp + + \right]} + \vv_{\left[\perp - +\right]} )^2 ~&=&~(\vv_{\left[\perp + -\right]} + \vv_{\left[\perp - -\right]})^2\,, \\
(\vv_{\left[+ \perp +\right]} + \vv_{\left[- \perp + \right]} )^2 ~&=&~(\vv_{\left[+ \perp -\right]} + \vv_{\left[- \perp -\right]})^2\,.
\eea
%\bea
%p_{+++}\,p_{-++} + p_{+-+}\,p_{--+} - p_{+++}\,p_{+-+} - p_{-++}\,p_{--+} ~=~ \nonumber\\
%p_{++-}\,p_{-+-} + p_{+--}\,p_{---} - p_{++-}\,p_{+--} - p_{-+-}\,p_{---}\,. 
%\eea
Subtracting the second equality from the first will give us the equality that involves only the probabilities because all of the cross terms are canceled due to the angle relations:
\bea
&p_{\{+++\}}\,p_{\{-++\}} ~+~ p_{\{+-+\}}\,p_{\{--+\}} ~-~ p_{\{+++\}}\,p_{\{+-+\}} ~-~ p_{\{-++\}}\,p_{\{--+\}}  \nonumber \\
&~=~ p_{\{++-\}}\,p_{\{-+-\}} ~+~ p_{\{+--\}}\,p_{\{---\}} ~-~ p_{\{++-\}}\,p_{\{+--\}} ~-~ p_{\{-+-\}}\,p_{\{---\}}\,.
\eea
This equality implies the following relation:
\beq
\label{ac_bc}
v_a\, \vb_{ac} ~=~  v_b\, \vb_{bc}\,.
\eeq
Two more constraints can be obtained by considering the other two pairs of qubits.
\bea
(\vv_{\left[\perp + + \right]} + \vv_{\left[\perp + -\right]} )^2 - (\vv_{[++\perp]} + \vv_{[-+\perp]})^2 &=& (\vv_{\left[\perp - +\right]} + \vv_{\left[\perp - -\right]})^2 - (\vv_{[+-\perp]} + \vv_{[--\perp]})^2 \,,  \nonumber\\
(\vv_{\left[+ \perp +\right]} + \vv_{\left[+ \perp - \right]} )^2 - (\vv_{[++\perp]} + \vv_{[+-\perp]})^2  &=& (\vv_{\left[- \perp + \right]} + \vv_{\left[- \perp -\right]})^2 - (\vv_{[-+\perp]} + \vv_{[--\perp]})^2 \,. \nonumber
\eea
The cross terms are again cancelled and we obtain two other relations of the probabilities. They will require the two other relations between the invariants:
\beq
\label{ab_bc}
v_a\,\vb_{ab} = v_c\, \vb_{bc}\,, ~\quad~ v_b\,\vb_{ab} = v_c\,\vb_{ac}\,.
\eeq
These relations suggest that all of the invariants of two qubits are related:
\beq
\label{vII_relation}
\la \vv_a\vv_b \VEC V_{ab} \ra = \la \vv_a\vv_c \VEC V_{ac} \ra  = \la \vv_b\vv_c \VEC V_{bc} \ra ~\equiv~ \vbb2\,.
\eeq
Naturally this means:
\beq
\vb_{ab} = \frac{\vbb2}{v_a\, v_b}\,,~\quad~ \vb_{bc} = \frac{\vbb2}{v_b\, v_c}\,, ~\quad~  \vb_{ac} = \frac{\vbb2}{v_a\, v_c}\,.
\eeq
We can also define:
\beq
\vbb3 ~\equiv~ \la \vv_a \vv_b \vv_c\, \VEC V_{abc} \ra = v_a\, v_b\, v_c\, \vb_{abc}\,.
\eeq
Therefore, the probabilities are parametrized by the five invariants: ($v_a\,, v_b\,, v_c\,, \vbb2\,, \vbb3$).

The next step is to solve the vector sum equations in order to constrained or fixed the directions of the twelve vectors. To write down the equations specifically, we will use the vectors, ${\hvf ++}$, ${\hvs ++}$ and ${\hvth ++}$ as the references and specify the directions of the other vectors by the angles they make with these references. We denote their angles by:
\bea
{\hvf +-} ~\rightarrow~ \phi_a^{(1)}\,,~\quad~ &{\hvf -+} ~\rightarrow~ \phi_a^{(2)}\,,& ~\quad~ {\hvf --} ~\rightarrow~ \phi_a^{(3)}\,, \\
{\hvs +-} ~\rightarrow~ \phi_b^{(1)}\,,~\quad~ &{\hvs -+} ~\rightarrow~ \phi_b^{(2)}\,,& ~\quad~ {\hvs --} ~\rightarrow~ \phi_b^{(3)}\,, \\
{\hvth +-} ~\rightarrow~ \phi_c^{(1)}\,,~\quad~ &{\hvth -+} ~\rightarrow~ \phi_c^{(2)}\,,& ~\quad~ {\hvth --} ~\rightarrow~ \phi_c^{(3)}.
\eea
Due to the pure state condition, these angles are related by \eqref{angle_cond}. We can define the following six angles in order to incorporate the relations.
\bea
\label{angle_relation}
\phi_a^{(2)} = \phi_b^{(2)} \equiv \phi_{ab} \,,& \phi_a^{(3)} - \phi_a^{(1)} = \phi_b^{(3)} - \phi_b^{(1)} \equiv \phi^{\prime}_{ab}\,, \nonumber \\
\phi_a^{(1)} = \phi_c^{(2)} \equiv \phi_{ac} \,,& \phi_a^{(3)} - \phi_a^{(2)} = \phi_c^{(3)} - \phi_c^{(1)} \equiv \phi^{\prime}_{ac}\,, \nonumber \\
\phi_b^{(1)} = \phi_c^{(1)} \equiv \phi_{bc} \,,& \phi_b^{(3)} - \phi_b^{(2)} = \phi_c^{(3)} - \phi_c^{(2)} \equiv \phi^{\prime}_{bc}\,. 
\eea
Note that these six angles are the angles between vectors and will not be affected by any local rotation. Furthermore, out of these six angles, only five of them are independent.  We can expand the three vector sum equations \eqref{vector_sum}  to the six equations in components and express them with the above six angles. Even though the number of the equations is more than the number of the independent valuables, the solutions still exist. That is because the lengths of the vectors are not completely random but constrained by \eqref{vII_relation}. Furthermore, the solutions of the vector sum equations come in pairs in general. If we flip the signs of all angles in one solution, the result is still a solution. The relation between these two solutions is similar to a conjugate pair (i.e., the coefficients of one state are complex conjugate to the other). Naturally, if all of the angles in one solution are either $0$ or $\pi$, then its pair is itself. Moreover, there are some special states in which all of the twelve vectors vanishing. We will talk about them later.

Unfortunately, we are unable to solve the vector sum equations analytically to obtain the expression of the six angles in terms of the five invariants. However, we have solved the equations numerically for many random combinations of the invariants. We find out that if the solutions exist at all, we get only one conjugate pair in general. In some special cases, we get only one solution with all of the angles being either $0$ or $\pi$. This suggests that the vector sum equations are enough to fix all of the six angles (up to the overall sign) in general. Therefore, the directions of all of the twelve vectors are basically determined once the directions of the three references are fixed. Since those directions can be changed by the local axial rotations of the three qubits, the degrees of freedom associated with them are actually local. Therefore, we can conclude that the only non-local (continuous) degrees of freedom in the density operator are coming from the five invariants.
 
% Note that the directions of the twelve vectors are basically determined by the six angles and the directions of the references. The angles are fixed by the vector sum equations and the directions of the references can be changed by the local axial rotations. This implies that the degrees of freedom left in the directions of the twelve vectors are local after we fix the five invariants and this is also true for the directions of all $\VEC X$-terms.

We can also count the number of the degrees of freedom to verify this. A three-qubit pure state is determined by the sixteen parameters (eight complex numbers). After applying the normalization condition and removing the global phase, we left with the fourteen parameters. Six of them are related to the directions of $\hv_a$, $\hv_b$ and $\hv_c$. Three\footnote{Note that unlike the two-qubit case, all of the three local axial rotations are independent degrees of freedom.} of them are related to the local rotations around the spins. Therefore, we left with the five invariants which are the only non-local degrees of freedom.

\subsubsection{The relations with the invariants in papers}

We have shown how to construct the five invariants and they are the only non-local degrees of freedom that we need to describe a general pure state. However, we define the five invariants in a natural way of the MSTA formulation and they are different from the invariants generally used in papers. In this section, we would like to find out the relation between our formulation of the invariants and the others. Specifically, we will use the invariants defined in \cite{sudbery2001local} as our reference.

There are totally six invariants defined in \cite{sudbery2001local}: $(I_1,I_2,...,I_6)$. The first one, $I_1$, is related to the norm of the state. Since we are dealing with the density operator with the unit trace, $I_1$ will not appear in our formulation. The next three invariants are defined in the matrix formulation by:
\beq
I_2 = \Tr(\rho_a^2)\,, ~\quad~ I_3 = \Tr(\rho_b^2)\,,  ~\quad~ I_4 = \Tr(\rho_c^2)\,, 
\eeq
where $\rho_a$, $\rho_b$ and $\rho_c$ are the reduced density operators obtained by tracing out the other qubits. The relations of these invariants with the values $v_a$, $v_b$ and $v_c$ are clear:
\beq
I_2 \rightarrow 2\left\la\left( \frac{1+\vv_a}{2} \right)^2 \right\ra ~=~ \frac{1+v_a^2}{2}\,, ~\quad~ 
I_3 \rightarrow  \frac{1+v_b^2}{2}\,, ~\quad~ I_4 \rightarrow  \frac{1+v_c^2}{2}\,. 
\eeq

The next one is defined by:
\bea
I_5 &= 3 \Tr[ (\rho_a \otimes \rho_b) \rho_{ab} ] - \Tr(\rho_a^3) - \Tr(\rho_b^3)\,, \\
&= 3 \Tr[ (\rho_a \otimes \rho_c) \rho_{ac} ] - \Tr(\rho_a^3) - \Tr(\rho_c^3)\,, \\
&= 3 \Tr[ (\rho_b \otimes \rho_c) \rho_{bc} ] - \Tr(\rho_b^3) - \Tr(\rho_c^3)\,.
\eea
The first expression is translated to:
\bea
I_5 &\rightarrow&  \left\la 12\, \left(\frac{1 + \vv_a}{2} \right) \left(\frac{1+\vv_b }{2} \right) \left(\frac{1+\vv_a + \vv_b + \VEC V_{ab}}{4} \right) - 2\left( \frac{1+\vv_a}{2} \right)^3  - 2\left( \frac{1+\vv_b}{2} \right)^3  \right\ra \nonumber  \\
&=& \frac{1}{4}\,(1 + 3 \la \vv_a\vv_b \VEC V_{ab} \ra ) =  \frac{1}{4}\,(1 + 3\, \vbb2 )\,.
\eea
The other two expressions of $I_5$ yield to $ \frac{1}{4}\,(1 + 3 \la \vv_a\vv_c \VEC V_{ac} \ra )$ and $ \frac{1}{4}\,(1 + 3 \la \vv_b\vv_c \VEC V_{bc} \ra )$ which should be all equivalent. This agree with the relation in \eqref{vII_relation}. So this conclude that $I_5$ has a very simple relation with $\vbb2$.

The last one is related to the 3-tangle \cite{coffman2000distributed} which is suggested to be a measurement of the three-way entanglement between qubits. It is more difficult to find out this invariant in our formulation since the 3-tangle is defined by the hyperdeterminant of the coefficients of the states. However, the canonical coordinate defined in \cite{sudbery2001local} is quite similar to our approach. Therefore we can translate the expression of $I_6$ in the section 5 of \cite{sudbery2001local} more directly. The result is:
\bea
I_6 &\equiv& \tau_{abc}^2 ~=~ (1+v_a^2-v_b^2-v_c^2)^2- 4\,\left(v_a-\frac{\vbb2}{v_a}\right)^2   \nonumber\\
&& - 4\,( (v_b+v_c)\,\vv_{\left[\perp,+,+\right]} + (v_b-v_c)\,\vv_{\left[\perp,+,-\right]}  + (v_c-v_b)\,\vv_{\left[\perp,-,+\right]}  - (v_b+v_c)\,\vv_{\left[\perp,-,-\right]} )^2\,, \nonumber
\eea
where $\tau_{abc}$ denotes the 3-tangle. Clearly, this formula can only be used after the vector sum equations are solved. It is much better if we can find a more direct relation to the five invariants of our formulation. Fortunately, such relation indeed exists:
\beq
\label{I6s}
I_6 = 1 - 2\,(v_a^2+v_b^2+v_c^2)  - 2\,( v_a^2\,v_b^2 + \,v_a^2\,v_c^2 + v_b^2\,v_c^2)+ (v_a^4 + v_b^4 + v_c^4) + 4\,(\vbb2+\vbb3)\,.
\eeq
%= (1+v_1-v_2-v_3)  (1-v_1+v_2-v_3)  (1-v_1-v_2+v_3)  (1+v_1+v_2+v_3) +4(\vbb2+\vbb3 - 2\,v_1 v_2 v_3)
This formula is found out by studying several examples and verified by many random states. It may be possible to derive this formula in a parallel way as the original paper \cite{coffman2000distributed} but we will leave this problem open for now.

\subsubsection{The ranges of the five invariants and the special states}

Before we can use the five invariants to describe the non-local part of the state space, we need to know the physical ranges of the five invariants (i.e., where the state exist). There are several conditions to ensure the state is physical. The most obvious condition is that all of the eight probabilities in \eqref{p_cc} must be nonnegative. This condition imposes a strong constraint on the ranges of the five invariants and ensures that the lengths of the twelve vectors are real and nonnegative. Furthermore, from the reduced density operators, we also need to require that the three invariants $v_a$, $v_b$ and $v_c$ are between $0$ and $1$.

The last condition is that we need to ensure the solutions of the vector sum equations \eqref{vector_sum} actually exist. To study this condition, we can check what happen on the boundary between the part that the solutions exist and the other part that do not. We found out that on the boundary, all of the vectors of each qubit are either vanishing or lie on a line. This can be understood in the following way. We can picture the set of four vectors of each qubit in a solution form a quadrilateral. The shapes of the three quadrilateral keep changing as we change the lengths of the vectors by tuning the five invariants. The solution become unavailable when all of the vectors lie on a line and we further change the lengths of vectors in certain way. For example, we may have the following equality satisfied on some point of the boundary:
\beq
v_{[\perp++]} - v_{[\perp+-]} - v_{[\perp-+]} - v_{[\perp--]} ~=~ 0\,,
\eeq
where $v_{[...]}$ is just the length of the vector $\vv_{[...]}$. Then the solution is quite clear that all of the vectors should lie on a line with $\vv_{[\perp++]}$ pointing at the opposite direction of the other vectors. However, if we further change the invariants such that $v_{[\perp++]} $ increase while the others decrease, the solution will no longer exist. Therefore, it is reasonable to assume the boundary located at the points where the following function is equal to zero:
\beq
\mathcal{F}_a \equiv \prod_{s_1,s_2,s_3=+,-}\,({\svf ++} + s_1\, {\svf +-} + s_2\,  {\svf -+} + s_3\, {\svf --} )\,.
\eeq
This function is the product of the eight terms with the different combinations of the lengths of the four vectors of the qubit $a$. Interestingly, the same function can be obtained by using the vectors of the other two qubits. It can be expressed in terms of the five invariants:
\beq
\label{functionF}
\mathcal{F} \equiv \mathcal{F}_a = \mathcal{F}_b = \mathcal{F}_c =\frac{(v_a^2 v_b^2 v_c^2 -  \vbb2 \vbb3)^2}{4096\,v_a^6\,v_b^6\,v_c^6}\, \mathcal{B}(v_a,v_b,v_c,\vbb2,\vbb3)\,,
\eeq
where $\mathcal{B}$ is:
\bea
\label{Bdef}
\mathcal{B}(v_a,v_b,v_c,\vbb2,\vbb3) =& -{\vbb 3}^3 + (\beta + \vbb2)\,\vbb3^2 + (\alpha\, \vbb2^2 - 2\, \beta\, \vbb2 + \gamma\,(1-\alpha))\,\vbb3 \\
& + \vbb2^4 - \alpha\, \vbb2^3 + (\beta - 2\,\gamma)\,\vbb2^2 - \gamma\, (1 - \alpha)\,\vbb2 + \gamma^2\,,
\eea
with $\alpha$, $\beta$ and $\gamma$:
\beq
\alpha \equiv v_a^2 + v_b^2 + v_c^2\,, ~\quad~ \beta \equiv v_a^2\,v_b^2 + v_a^2\,v_c^2 + v_b^2\,v_c^2\,,  ~\quad~ \gamma \equiv v_a^2\, v_b^2\, v_c^2\,. 
\eeq

The function $\mathcal{F}$ is zero when $\vbb2\,\vbb3$ is equal to $v_a^2 v_b^2 v_c^2$ or when $\mathcal{B}$ is zero. At first, we examine the first possibiity. If $\vbb2\,\vbb3$ is equal to $v_a^2 v_b^2 v_c^2$, we have:
\bea
&{\svf ++}= {\svf--}\,,~\quad~ {\svf +-}= {\svf-+}\,, ~\quad~  {\svs ++}= {\svs--}\,, ~\quad~  {\svs +-}= {\svs-+} \nonumber \\
 &{\svth ++}= {\svth--}\,, ~\quad~  {\svth +-}= {\svth-+}\,.
\eea
Because of the above relations and the angle relations in \eqref{angle_cond}, the solution can not have all of the vectors lie on a line in general. Therefore, the boundary should be located at $\mathcal{B}=0$.

By checking several examples, we found out that the solutions exist at the points where $\mathcal{B}$ is non-positive. Therefore, we can summary all of the constraints on the invariants as the following:
\begin{empheq}[left=\empheqlbrace]{align}
& p_{\{i j k\}} \geq 0\,, ~\quad~ i, j, k = \pm\,, \label{ppc}\\
& 0 \leq v_a \leq 1\,,  \quad 0 \leq v_b \leq 1\,, \quad 0 \leq v_c \leq 1\,, \\
& \mathcal{B}(v_a,v_b,v_c,\vbb2,\vbb3)  \leq 0 \label{Bzero}\,.
\end{empheq}
The states basically exist inside of a cube of the ($v_a, v_b, v_c$) space. Therefore, the main task is to study the other two constraints on the two values $\vbb2$ and $\vbb3$ given a point inside of that cube. We will check the allowed regions in the $(\vbb2,\vbb3)$ graph for several combinations of $v_a$, $v_b$ and $v_c$ to show the general traits. However, before we do that, we will study several special states with some particular values of the $\vbb2$ and $\vbb3$. 

\begin{paragraph}{The seed states ($\vbb2=\vbb3=v_a\,v_b\,v_c$):}
this is a very special case when four of the probabilities vanish. The remaining ones are:
\bea
p_{\{+++\}}=\frac{1}{4}\,(1 + v_a + v_b + v_c)\,, ~&\quad~ p_{\{+--\}}=\frac{1}{4}\,(1 + v_a - v_b - v_c)\,, \\
p_{\{-+-\}}=\frac{1}{4}\,(1 - v_a + v_b - v_c)\,, ~&\quad~ p_{\{--+\}}=\frac{1}{4}\,(1 - v_a - v_b + v_c)\,.
\eea
The nonnegativity of these probabilities impose an additional constraint on $v_a$, $v_b$ and $v_c$:
\beq
1 + 2\,v_{min} \geq v_a + v_b + v_c\,, ~\quad~ v_{min} \equiv \min\,(v_a,v_b,v_c)\,.
\eeq
Since $\mathcal{B}$ is exactly zero, this is the only additional condition for the existence of the state in this case and these states are located on the boundary of the allowed region in the ($\vbb2,\vbb3$) graph. However, notice that all of the twelve vectors vanish and therefore we will not be able to obtain any information from the vector sum equations. How do we deal with the remaining degrees of freedom in the directions of the $\VEC X$-terms? In fact, we can show that all of the remaining degrees of freedom are just local. At first, notice that we only have three independent directions needed to be determined (i.e., the other three are determined by the pure state condition). These three directions are controlled by the local axial rotations of the three qubits. For example, if we choose the directions of the following $\VEC X$'s as the independent ones:
\beq
\VEC X_{\{+++,+--\}}\,,  ~\quad~ \VEC X_{\{+--,-+-\}}\,, ~\quad~ \VEC X_{\{-+-,--+\}}\,.
\eeq
Then we can change the azimuthal angles of the above $\VEC X$'s by $\phi_1$, $\phi_2$ and $\phi_3$ respectively with the local rotations around the spins of the three qubits by the following angles:
\beq
\phi_a=\frac{1}{2}\,(\phi_1+2\phi_2+\phi_3)\,, \quad \phi_b=\frac{1}{2}\,(\phi_1+\phi_3)\,, \quad \phi_c=\frac{1}{2}\,(\phi_1-\phi_3).
\eeq
Therefore, we can confirm that the degrees of freedom in the directions of $\VEC X$'s are local.

This special case is quite important because it provide a further constraint on the values $v_a$, $v_b$ and $v_c$. The reason is that if the state in this case does not exist for a particular combination of $(v_a,v_b,v_c)$, then there will be no other state (i.e., the state with other values of $\vbb2$ and $\vbb3$) exist for such combination. For example, if we have $v_c=1$ but $v_a \neq v_b$, the state in this case will not exist and there will be no state exist at all. More generally, if we have the values $v_a$, $v_b$ and $v_c$ inside of the cube and satisfy:
\beq
0\leq v_a \leq v_b \leq v_c \leq 1\,, ~\quad~   1+v_a < v_b + v_c\,, \nonumber
\eeq
then we can show that it is not possible to find any combination of $(\vbb2, \vbb3)$ such that all of the following conditions are satisfied:
\beq
p_{\{+--\}} \geq 0\,, ~\quad~ p_{\{-++\}} \geq 0\,, ~\quad~ p_{\{---\}} \geq 0\,. \nonumber
\eeq
Due to this special property, we will refer to the state in this case as the seed state. Their existence condition provide the additional criterion for the allowed values of the $v_a$, $v_b$ and $v_c$.

This case includes several special states:
\begin{itemize}
\item The product states: when $v_a=v_b=v_c=1$, the state is $\{+++\}$\,.
\item The bi-separable states: when one of the $v$ equal to one. For example if we have $v_c$ equal to one, the constraint require the other two $v$'s to be the same value (i.e., $v_a=v_b=v$). If this value is smaller than one, the state is bi-separable and that is when only the first two qubits are entangled. The density operator $\rho_{abc}$ can be written as the product of $\rho_{ab}$ and $(1+\hv_c)/2$, where $\rho_{ab}$ is some general two qubit state defined in \eqref{rho12_red}.
\end{itemize}

Using the formula \eqref{I6s}, we can check the invariant $I_6$ related to the 3-tangle for the seed states:\beq
\label{I6seed}
I_6^{(s)} ~=~ (1 + v_a - v_b - v_c) (1 - v_a + v_b - v_c) (1 - v_a - v_b + v_c) (1 + v_a + v_b + v_c)\,.
\eeq
From the formula, we can verify that the $I_6$ for both of the product states and the bi-separable states are zero. This is to be expected since the 3-tangle is supposed to measure the three-way entanglement between qubits.

\end{paragraph}

\begin{paragraph}{The negative seed states ($\vbb2=\vbb3= - v_a\,v_b\,v_c$):}
this is also a very special case quite similar to the previous one. The four of the probabilities vanishing and the remaining ones are:
\bea
p_{\{++-\}}=\frac{1}{4}\,(1 + v_a + v_b - v_c)\,, ~&\quad~ p_{\{+-+\}}=\frac{1}{4}\,(1 + v_a - v_b + v_c)\,, \\
p_{\{-++\}}=\frac{1}{4}\,(1 - v_a + v_b + v_c)\,, ~&\quad~ p_{\{---\}}=\frac{1}{4}\,(1 - v_a - v_b - v_c)\,.
\eea
The nonnegativity of the probabilities require: 
\beq
\label{cond_nss}
v_a+v_b+v_c \leq 1\,.
\eeq
Again, since $\mathcal B$ is exactly zero, this is the only additional existence condition and the states in this case are located on the boundary of the allowed region in the ($\vbb2,\vbb3$) graph. Furthermore, just like the previous case, all of the twelve vectors vanish and the remaining degrees of freedom in the $\VEC X$-terms are local. Because of their similarity, we will refer to the states in this case as the \emph{negative} seed states.

This case includes several well-known special states:
\begin{itemize}
\item The W-states: $v_a=v_b=v_c=\frac{1}{3}$. Up to some local rotations, the density operator can be written as:
\bea
\rho_{W} &=& \frac{1}{8}\,\left(1+ \frac{1}{3}\,(\hv_a + \hv_b + \hv_c) - \frac{1}{3}\,(\hv_a \hv_b + \hv_a \hv_c + \hv_b \hv_c) - \hv_a \hv_b \hv_c   \nonumber \right. \\
&&+ \frac{2}{3}\,\hv_{a\perp}\hv_{b\perp} (1- \hv_a \hv_b)(1+\hv_c) + \frac{2}{3}\,\hv_{a\perp}\hv_{c\perp} (1- \hv_a \hv_c)(1+\hv_b) \nonumber \\
&&\left. +\frac{2}{3}\, \hv_{b\perp}\hv_{c\perp} (1- \hv_b \hv_c)(1+\hv_a) \right)\,, 
\eea
where $\hv_{a\perp}$, $\hv_{b\perp}$ and $\hv_{c\perp}$ are some unit vectors orthogonal to $\hv_a$, $\hv_b$ and $\hv_c$ respectively.
\item The generalized W-states: $v_a+v_b+v_c=1$. The structure of the density operator is similar to the previous one with only some of the coefficients changed.
\end{itemize}
% &=& \frac{1}{3}\,(\{++-\} + \{+-+\} + \{-++\}) + \frac{1}{3}\,( \VEC X_{\{++-,+-+\}} + \VEC X_{\{+-+,-++\}} +  \VEC X_{\{+-+,-++\}}) \nonumber \\

The $I_6$ of the negative seed state is:
\beq
\label{I6nseed}
I_6^{(ns)} ~=~ (1 - v_a + v_b + v_c) (1 + v_a - v_b + v_c) (1 + v_a + v_b - v_c) (1 - v_a - v_b - v_c)\,.
\eeq
From the formula, we can clearly see that the (generalized) W-states have zero 3-tangle.

\end{paragraph}

\begin{paragraph}{The maximum 3-tangle states:} given a combination of $(v_a,v_b,v_c)$, the state in this case has the maximum 3-tangle if it exist. The values $\vbb2$ and $\vbb3$ in this case should be set to:
\beq
\vbb2 = v_{min}^2\,,~\quad~ \vbb3=\frac{v_a^2\,v_b^2\,v_c^2}{v_{min}^2}\,.
\eeq
The nonnegativity of the probabilities require:
\beq
\label{m3tan_cond}
v_{min}^2 \geq v_a\,v_b\,v_c\,.
\eeq
Since $\mathcal{B}$ is also zero, this is the only additional condition for the existence of the states in this case. The states in this case are also on the boundary just like the (negative) seed states. However, the twelve vectors do not vanish in general but lie on a line.

The $I_6$ of the state in this case is:
\beq
\label{I6m}
I_6^{(m)} = (1-v_a^2-v_b^2-v_c^2 + 2\, v_{min}^2)^2\,.
\eeq
%\bea
%I_6 &=~ (1 + v_a^2 - v_b^2 - v_c^2)^2 \,, ~\quad~ \text{if} ~ v_a \leq v_b\,,~v_a \leq v_c  ~\text{and}~  v_a \geq v_b v_c\,, \nonumber \\
%      &=~ (1 - v_a^2 + v_b^2 - v_c^2)^2 \,, ~\quad~ \text{if} ~ v_b \leq v_a\,,~ v_b \leq v_c ~\text{and}~ v_b \geq v_a v_c\,, \nonumber\\
%      &=~ (1 - v_a^2 - v_b^2 + v_c^2)^2 \,, ~\quad~ \text{if} ~ v_c \leq v_a\,,~ v_c \leq v_b ~\text{and}~ v_c \geq v_a v_b \,.
%\eea
However, if the values of $v_a$, $v_b$ and $v_c$ does not satisfy the existence condition in \eqref{m3tan_cond}, then the seed state should be the one has the maximum 3-tangle in that case. This can be seen by subtracting \eqref{I6seed} from \eqref{I6m}:
\beq
I_6^{(m)}-I_6^{(s)} = 4\,(v_{min}^2- v_a\, v_b\, v_c)^2/v_{min}^2\,.
\eeq
When $v_{min}^2=v_a\,v_b\,v_c$, their 3-tangles become identical. If $v_{min}^2<v_a\,v_b\,v_c$, the state in this case does not exist and the seed state become the state with the maximum 3-tangle. 

\end{paragraph}

\begin{paragraph}{The zero 3-tangle states:} given a combination of $(v_a,v_b,v_c)$, the state in this case has zero 3-tangle if it exist. This can be obtained by solving the following equations:
\beq
I_6~=~0\,, ~\quad~ \mathcal{B} ~=~0\,,
\eeq
where $I_6$ and $\mathcal{B}$ are defined in \eqref{I6s} and \eqref{Bdef}. We set $\mathcal{B}=0$ because we expect the zero 3-tangle states to be on the boundary. There are multiple solutions but only one of them is possible to have nonnegative probabilities in general and that is:
\bea
\label{zero_vIIvIII}
\vbb2 &=& -\frac{1-v_a^2-v_b^2-v_c^2}{2} + \xi\,, \nonumber\\
\vbb3 &=&\frac{1}{4}\,(1-v_a^4-v_b^4-v_c^4+2\,(v_a^2 v_b^2 + v_a^2 v_c^2 + v_b^2 v_c^2)) - \xi\,. 
\eea
%\frac{1}{4}\,(1+ (-v_1+v_2+v_3)(v_1-v_2+v_3)(v_1+v_2-v_3)(v_1+v_2+v_3))
where $\xi$ is:
\beq
\xi \equiv \sqrt{ \left( \frac{1+v_a^2-v_b^2-v_c^2}{2} \right)\left( \frac{1-v_a^2+v_b^2-v_c^2}{2} \right)\left( \frac{1 - v_a^2 - v_b^2 + v_c^2}{2} \right)}\,.
\eeq
The nonnegativity of the probabilities require:
\beq
1 + 2\,v_{min} \geq v_a + v_b + v_c \geq 1\,.
\eeq
The higher bound is just the existence condition of the seed state. The lower bound show that the zero 3-tangle states are compensating with the negative seed states. This can be seen by comparing the lower bound with the existence condition of the negative seed states in \eqref{cond_nss}. This means that the whole state space is divided to two part by the surface: $v_a+v_b+v_c=1$ and the negative seed states only exist at the part where $v_a+v_b+v_c\leq 1$ while the zero 3-tangle states only exist when $v_a+v_b+v_c\geq 1$. On the boundary, the states of two cases match with each others. This can be seen from \eqref{zero_vIIvIII} that when $v_a+v_b+v_c=1$, the $(\vbb2\,,\vbb3)$ reduced to $(-v_a v_b v_c, -v_a v_b v_c)$. Since the invariant $I_6$ is nonnegative, the state in this case has the minimum 3-tangle if it exist for a combination of $(v_a,v_b,v_c)$. If it does not exist, the negative seed state should have the minimum 3-tangle.

\end{paragraph}

\subsubsection{The $(\vbb2,\vbb3)$ graphs}

The two constraints \eqref{ppc} and \eqref{Bzero} are difficult to analyze in general. Therefore, we would like to show the allowed region in the ($\vbb2,\vbb3$) graph for several combinations of the values $v_a$, $v_b$ and $v_c$. By checking several examples, we can see the general traits of this region.

\begin{figure}
\begin{center}
\includegraphics[width=2in]{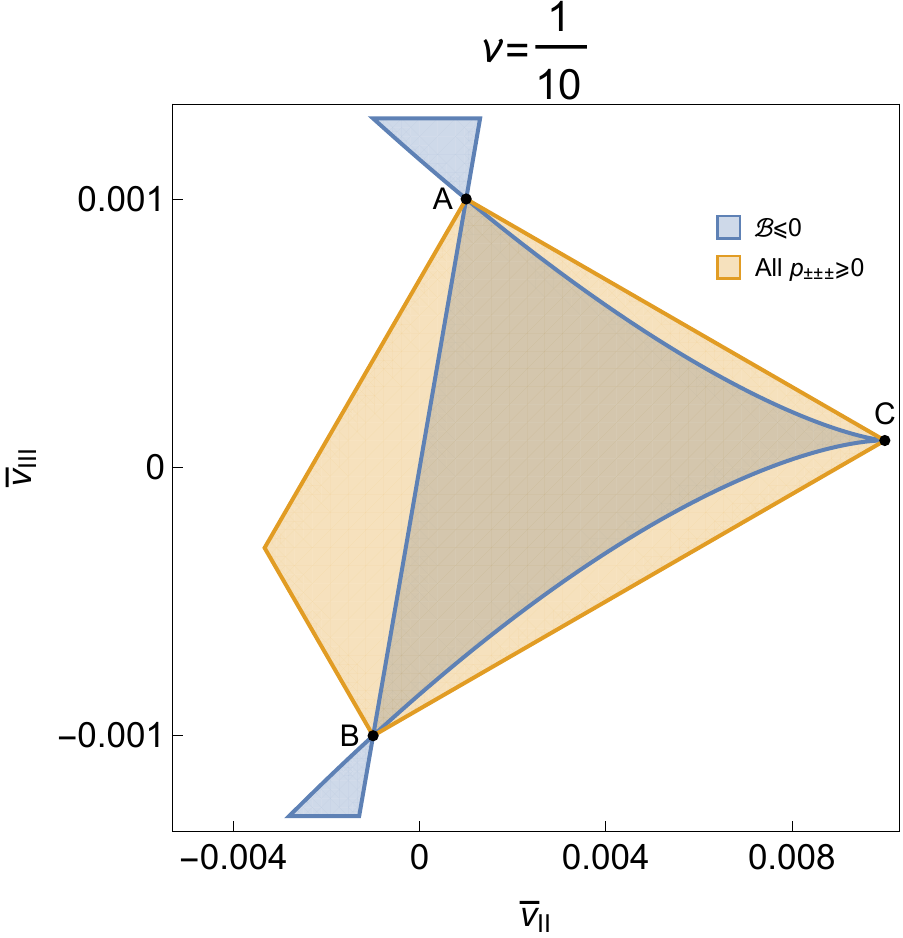}
\includegraphics[width=2in]{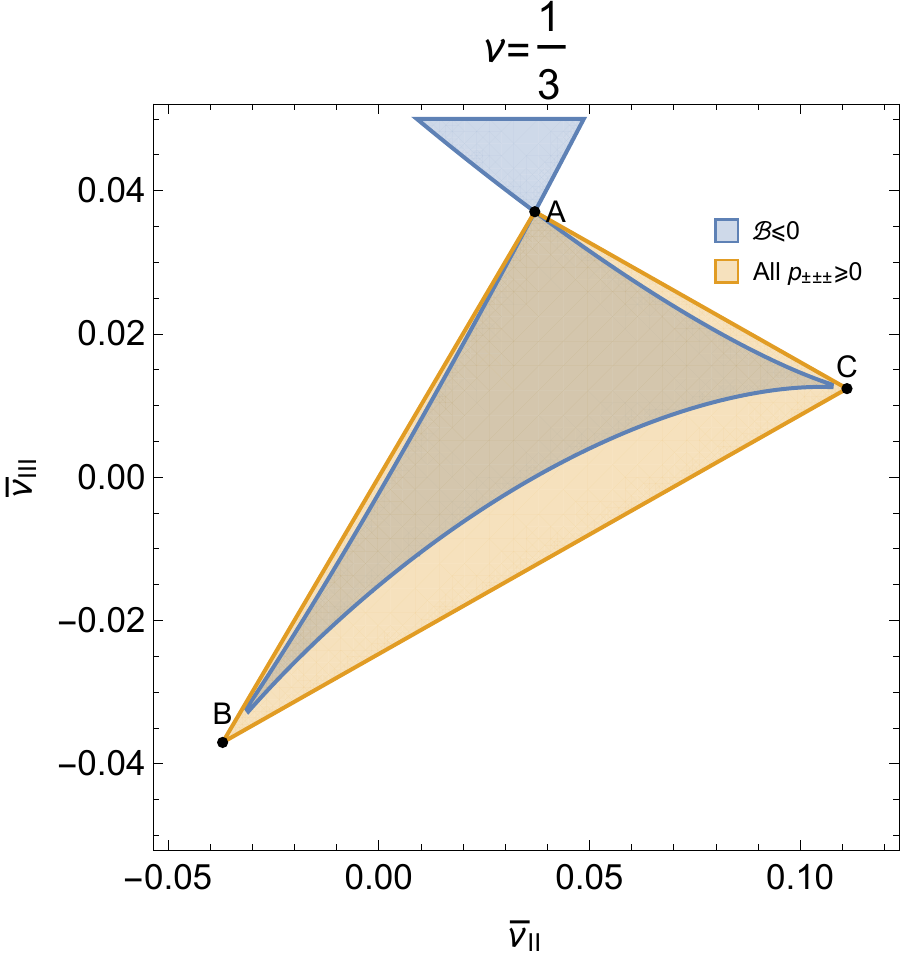}
\includegraphics[width=2in]{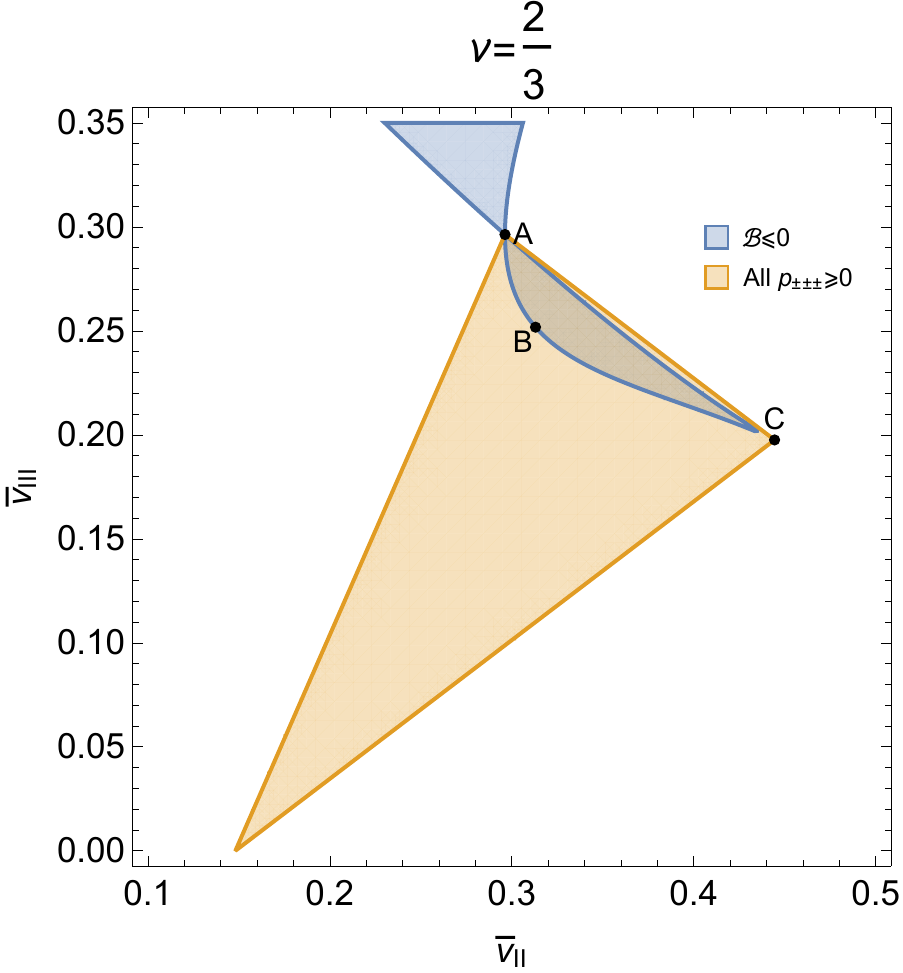}
\caption{The $(\vbb2,\vbb3)$ graphs for $v_a=v_b=v_c=v=\frac{1}{10}\,,\frac{1}{3}$ and $\frac{2}{3}$. The point A is the seed state located at $(\vbb2,\vbb3)=(v^3,v^3)$. The point C is the maximum 3-tangle state located at $(\vbb2,\vbb3)=(v^2,v^4)$. The point B is the minimum 3-tangle state and for $v\leq \frac{1}{3}$, it is the negative seed state located at $(\vbb2,\vbb3)=(-v^3,-v^3)$. For $v\geq\frac{1}{3}$, it indicate the zero 3-tangle state.}
\label{fig:rpvvv}
\end{center}
\end{figure}

At first, we consider some special examples when all of the three values are equal: $v_a=v_b=v_c=v$. When $v$ equal to zero, the vector part vanish and the states in that case will be discussed later. When $v$ equal to one, the state is just $\{+++\}$. Therefore, the value $v$ should be larger than zero and less than one. We pick up the three values $v=\frac{1}{10}\,,\frac{1}{3}$ and $\frac{2}{3}$ as our examples. We mark the regions where the conditions \eqref{ppc} and \eqref{Bzero} are satisfied separately in the ($\vbb2,\vbb3$) graph. The results are shown in fig. \ref{fig:rpvvv}. The graphs also indicate the three special states on the boundary: the seed state, the maximum 3-tangle state and the minimum 3-tangle state. The seed state is located at $(\vbb2,\vbb3)=(v^3,v^3)$. The maximum 3-tangle state is located at $(\vbb2,\vbb3)=(v^2,v^4)$. For $v \geq \frac{1}{3}$, the minimum 3-tangle state is the state with the zero 3-tangle:
\beq
(\vbb2\,,\vbb3) ~=~ \frac{1}{4}\,\left( -2 + 6\, v^2 + \sqrt{2}\,(1-v^2)^{3/2},\, 1 + 3\, v^4 - \sqrt{2}\,(1-v^2)^{3/2}\,. \right)
\eeq
However, when $v < \frac{1}{3}$, the zero 3-tangle state no longer exist. To find the minimum 3-tangle state, we can minimize the $I_6$:
\beq
I_6|_{v_1=v_2=v_3=v} ~=~ 1 - 6 v^2 - 3 v^4 + 4( \vbb2 + \vbb3)\,,
\eeq
under the constraint of \eqref{ppc} and \eqref{Bzero}. The minimum point is at $(\vbb2,\vbb3)=(-v^3,-v^3)$ which is the negative seed state as we have mentioned before.

\begin{figure}
\begin{center}
\includegraphics[width=2in]{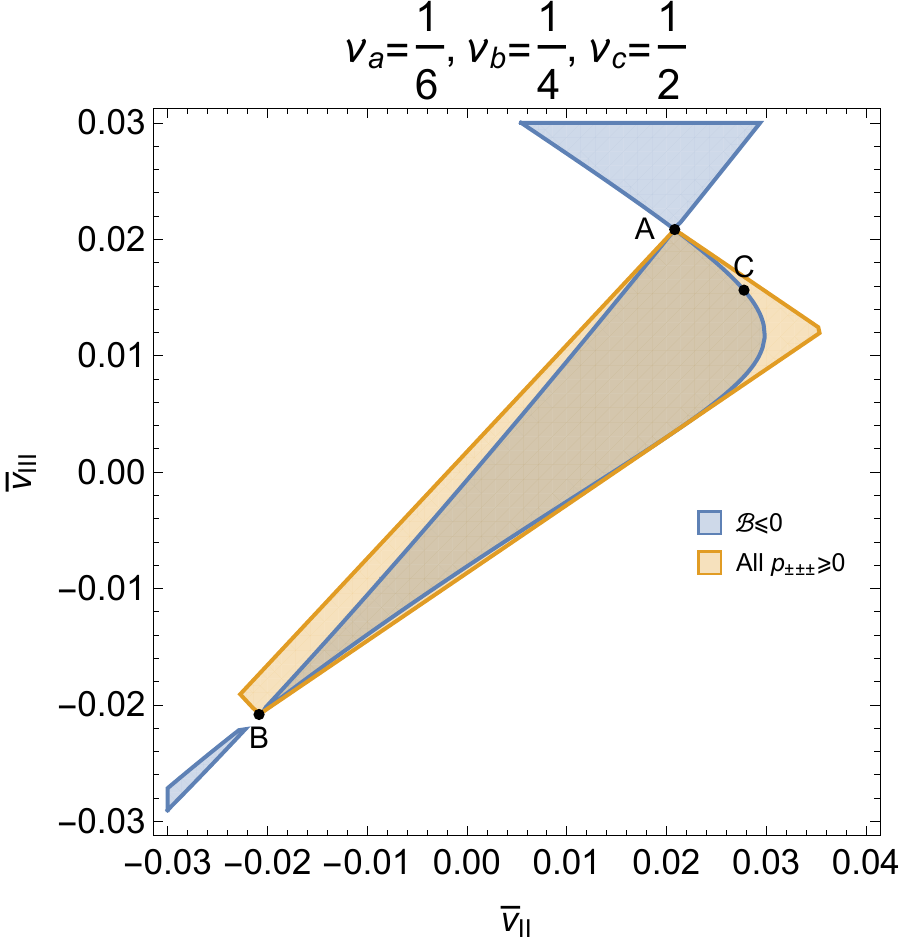}
\includegraphics[width=2in]{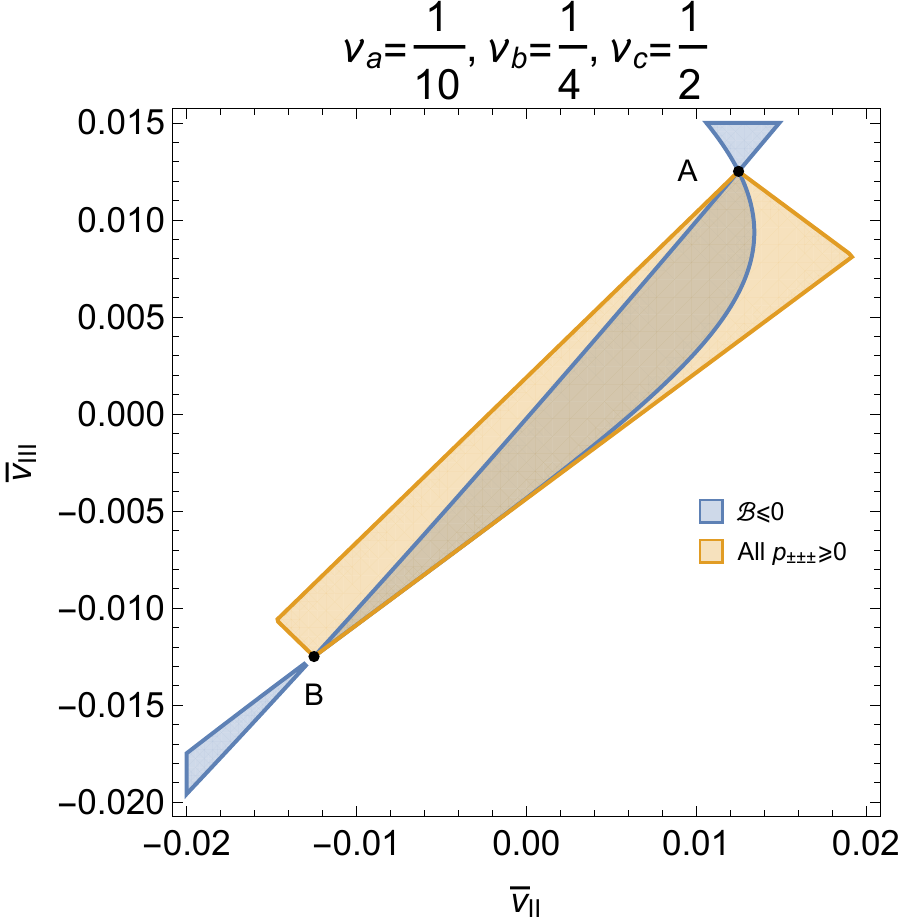}
\caption{The $(\vbb2,\vbb3)$ graphs for $(v_a,v_b,v_c)=(\frac{1}{6},\frac{1}{4},\frac{1}{2})$ and $(\frac{1}{10},\frac{1}{4},\frac{1}{2})$. In the first case, the maximum 3-tangle state exist and indicated by the point C. The point A is the seed state while the point B is the negative seed state.  }
\label{fig:rpvn}
\end{center}
\end{figure}

\begin{figure}
\begin{center}
\includegraphics[width=2in]{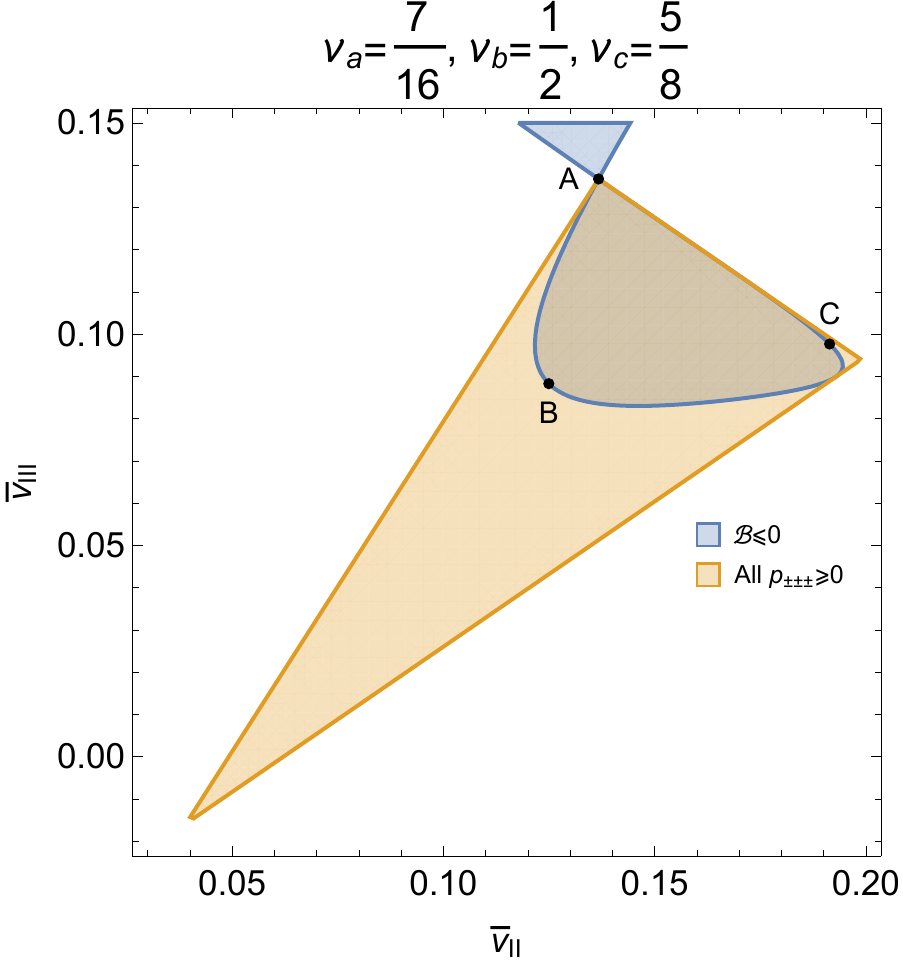}
\includegraphics[width=2in]{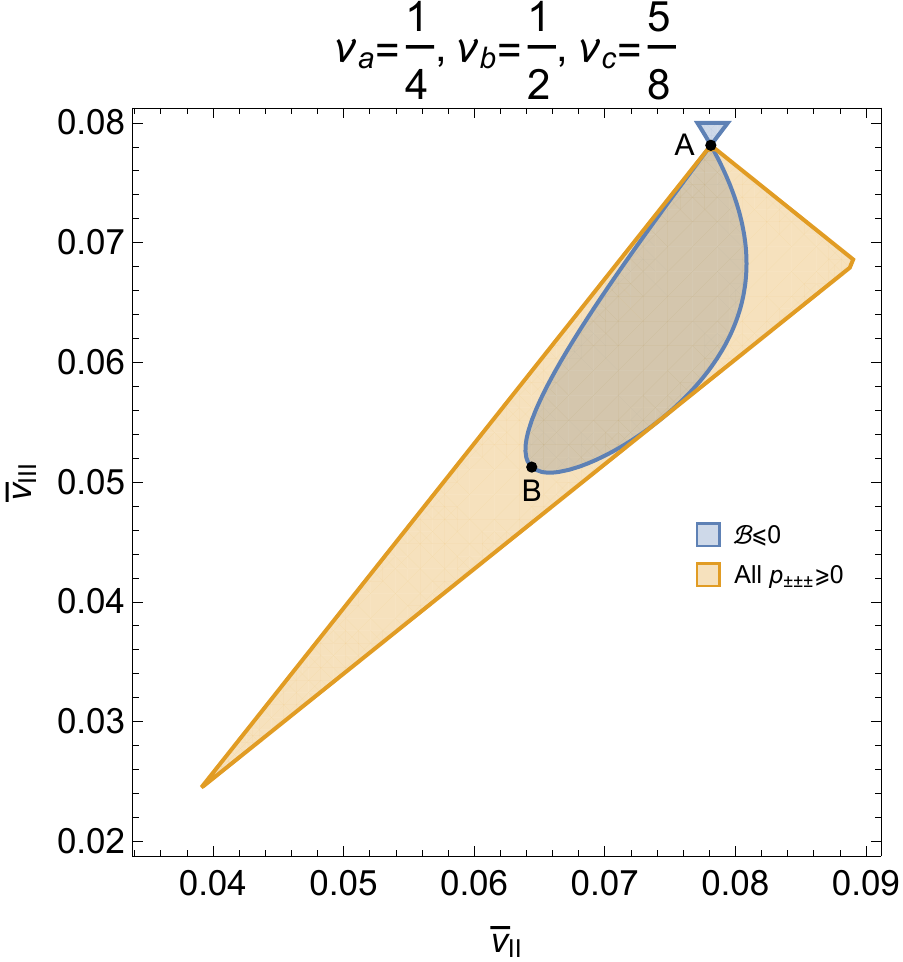}
\caption{The $(\vbb2,\vbb3)$ graphs for $(v_a,v_b,v_c)=(\frac{7}{16},\frac{1}{2},\frac{5}{8})$ and $(\frac{1}{4},\frac{1}{2},\frac{5}{8})$. In the first case, the maximum 3-tangle state exist and indicated by the point C. The point A is the seed state while the point B is the negative seed state. }
\label{fig:rpvz}
\end{center}
\end{figure}

We also pick two examples for the case when $v_a+v_b+v_c < 1$ and two other examples for the case when $v_a+v_b+v_c > 1$. The graphs are shown in fig. \ref{fig:rpvn} and fig. \ref{fig:rpvz}. From these graphs, we see that the states exist in a simply connected region of the $(\vbb2,\vbb3)$ graph surround by several special states. If we ignoring the local degrees of freedom, for now, there is only one state on the boundary of the region but two solutions inside (i.e., the conjugate pair with the opposite signs of the angles). Therefore, the state space in the ($\vbb2,\vbb3$) graph can be regarded as a sphere topologically. Furthermore, on the boundary of the ($v_a,v_b,v_c$) space, where the existence condition of the seed state is saturated (i.e., $1+2v_{min}=v_a+v_b+v_c$), we can show that only the seed state can satisfy the probability condition \eqref{ppc}. Therefore, the state space shrink to a point on the \emph{outside} boundary of the ($v_a,v_b,v_c$) space. However, we also have the \emph{inside} boundary where one or more of the values $v_a$, $v_b$ and $v_c$ go to zero. We will study this case in the following.

\subsubsection{The states with some of the vectors vanishing}

For the states with some of the three vectors (i.e., $\vv_a$, $\vv_b$ and $\vv_c$) vanishing, there will be some ambiguity for choosing the expansion product states. The expansion probabilities are no longer invariants if we just pick some random directions for the missing vectors. Even though the procedure to generate the local invariant labels no longer work, we can still obtain some information about the structure of the density operator from it for some cases.

On the other hand, notice that both of the invariants $\vbb2$ and $\vbb3$ become zero if any of the three vectors vanishes. By continuity, this means the state space shrink to a point on the \emph{inside} boundary of the ($v_a,v_b,v_c$) space just like the \emph{outside} boundary. This implies that if our description of the state space is consistent, the only degrees of freedom left in the density operator, in this case, should be just local besides of the remaining invariants. Therefore, we can take the limit of some special states (e.g. the seed states) to obtain the density operator in this case. The result should be able to represent the state up to some local rotations. We will discuss the three cases based on the number of the remaining $v$'s.

\begin{paragraph}{The states with two of the vectors remaining:}

we will assume $v_a$ equal to zero. Since $\vv_a$ do not appear in the density operator, the choice of the direction of $\hv_a$ become ambiguous and the values $\vb_{ab}$, $\vb_{ac}$ and $\vb_{abc}$ are no longer invariants. However, from the consistent conditions in \eqref{ac_bc} and \eqref{ab_bc}, we obtain:
\beq
\vb_{bc} = 0\,, ~\quad~ \frac{\vb_{ab}}{v_c} = \frac{\vb_{ac}}{v_b} \equiv \vb_2\,.
\eeq
We can use the two invariants $v_b$ and $v_c$ and the two other parameters $\vb_2$ and $\vb_3\equiv\vb_{abc}$ to parametrize the probabilities. The two new parameters are related to the original invariants $\vbb2$ and $\vbb3$ (which are both vanishing) formally by:
\beq
\vbb2 = v_a\,v_b\,v_c\,\vb_2\,, ~\quad~ \vbb3 = v_a\,v_b\,v_c\,\vb_3\,.
\eeq
The function $\mathcal{F}$ in \eqref{functionF} can be rewritten with these new parameters:
\beq
\mathcal{F} = \frac{v_b^2\,v_c^2}{4096}\,(1-\vb_2\vb_3)^2\,(\vb_2 - \vb_3)^2\,.
\eeq
Considering the continuity of this function extended from the part with small $v_a$, we find out the part that play the role of the function $\mathcal{B}$ is $(\vb_2-\vb_3)^2$. This means the existence of the state require $\vb_2$ equal to $\vb_3$. Combining the nonnegativity of the probabilities, the ranges of the parameters are:
\beq
v_b+v_c\leq 1\,, \quad -1 \leq (\vb_2 = \vb_3 \equiv \vb) \leq 1\,.
\eeq
The value of $\vb$ can be changed by choosing the different direction of the qubit $a$ in the expansion product states or by some local rotations on the density operator. By changing the value of $\vb$ to (minus) one, only four of the expansion probabilities remain and the state is resemble to the (negative) seed state with vanishing $v_a$. Using the similar argument before, we can say the degrees of freedom coming from the directions of the $\VEC X$-terms are all local. This conclude that all of degrees of freedom left in the density operator are local besides of the remaining invariants $v_b$ and $v_c$.

On the other hand, we can also follow the appropriate seed state (i.e., with $v_b+v_c \leq 1$) by taking $v_a$ to zero. It can be explicitly verified that for the result density operator $\vb_2$ is equal to $\vb_3$ and that value is indeed controlled by the direction of $\hv_a$ used. The $I_6$ for the states in this case is:
\beq
I_6 = (1-v_c^2)^2 -2\,(1+v_c^2)\,v_b^2+v_b^4\,.
\eeq

\end{paragraph}

\begin{paragraph}{The states with only one of the vectors remaining:} 

we assume both $v_a$ and $v_b$ equal to zero. In this case, besides of $v_c$, all of the other parameters (i.e., $\vb_{ab}$, $\vb_{ac}$, $\vb_{bc}$ and $\vb_{abc}$) are no longer invariants. But, from the consistent conditions, we can still deduce that $\vb_{bc}$ and $\vb_{ac}$ should be zero. Therefore, we have one invariant $v_c$ and two parameters $\vb_{ab}$ and $\vb_{abc}$ to parametrize the probabilities. On the other hand, the function $\mathcal{F}$ is exactly zero, so we can not get any information from it. Since $\vb_{ab}$ and $\vb_{abc}$ are no longer invariants, their values can be changed by some local rotations. We find out that if we change the values of $\vb_{ab}$ and $\vb_{abc}$ to the following:
\beq
(\vb_{ab}\,,\vb_{abc})= (v_c\,, 1)\,,
\eeq
only four of the expansion probabilities remain which are exactly the same as the seed state's with $v_a$ and $v_b$ vanishing. Therefore, we can conclude that the degrees of freedom associated to the directions of the $\VEC X$'s part is also completely local for the states in this case and once the value $v_c$ is fixed, the density operator is completely determined up to some local rotations.

If we want to write down the density operator, we can follow the seed state by taking the limit of $v_a$ and $v_b$ to zero. The result is:
\bea
\rho = \frac{1}{8}\,\Big( 1+ \vv_c + v_c\, \hv_a\hv_b +\hv_{a\perp}\hv_{b\perp}(1+v_c\, \hv_a \hv_b) + \sqrt{1-v_c^2}\,(\hv_{a\perp}\hv_{c\perp}+\hv_{b\perp}\hv_{c\perp})\nonumber\\
+ \sqrt{1-v_c^2}\,(\hv_{a\perp}\hv_{c\perp}+\hv_{b\perp}\hv_{c\perp})\, \hv_a\hv_b\hv_c  +(1+\hv_{a\perp}\hv_{b\perp})\, \hv_a\hv_b\hv_c  +  \hv_{a\perp}\hv_{b\perp}\vv_c\Big)\,,
\eea
where $\hv_a$ and $\hv_b$ in the products can be seen as coming from the limit when $\vv_a$ and $\vv_b$ are almost vanishing. We can see that the values $\vb_{ab}$ and $\vb_{abc}$ depending on the choice of the directions of the qubits $a$ and $b$ in the expansion product states. Moreover, the values can also be changed by rotating the direction of $\hv_a$ or $\hv_b$ in the density operator. The $I_6$ for the states in this case is:
\beq
I_6 = (1-v_c^2)^2
\eeq

\end{paragraph}

\begin{paragraph}{The states with none of the vectors remaining:} 

all of the invariants become zero. There is not much we can say about the states in this case by following the usual procedure. All of the remaining parameters are not invariants and can be changed by some local rotations. We will study the density operator in this case by taking the limit of several special states. At first, we follow the seed state by taking all of the values $v_a$, $v_b$ and $v_c$ to zero. In this limit, the density operator become:
\beq
\rho = \frac{1}{8}\,(1+\hv_{a\perp}\hv_{b\perp}+\hv_{a\perp}\hv_{b\perp}+\hv_{a\perp}\hv_{b\perp})(1+\hv_a\hv_b\hv_c)\,,
\eeq
where $\hv_{a\perp}$, $\hv_{b\perp}$ and $\hv_{c\perp}$ are some arbitrary unit vectors orthogonal to $\hv_a$, $\hv_b$ and $\hv_c$ respectively. If we follow the negative seed state, the density operator become:
\beq
\rho = \frac{1}{8}\,(1+\hv'_{a\perp}\hv'_{b\perp}+\hv'_{a\perp}\hv'_{b\perp}+\hv'_{a\perp}\hv'_{b\perp})(1-\hv_a\hv_b\hv_c)\,.
\eeq
On the other hand, we can also follow the maximum 3-tangle state by setting all of the values $v_a$, $v_b$ and $v_c$ equal to $v$ and then taking $v$ to zero. With this approach, we will obtain:
\beq
\rho= \frac{1}{8}\,(1+\hv_a\hv_b+\hv_a\hv_c+\hv_b\hv_c)(1+\hv''_{a\perp}\hv''_{b\perp}\hv''_{c\perp})\,.
\eeq
From the above density operators, it should be clear that they are all related to each other by some local rotations of three qubits. Furthermore, we can also see that they are all related to the GHZ state in \eqref{do_GHZ} by some local rotations. Therefore, we can conclude that the states for this case are the same with the GHZ state up to some local rotations. The $I_6$ for the states in this case is exactly one which is the maximum possible values for the 3-tangle.

\end{paragraph}

\subsubsection{The local degrees of freedom and the state space}

\begin{figure}
\begin{center}
\includegraphics[width=4in]{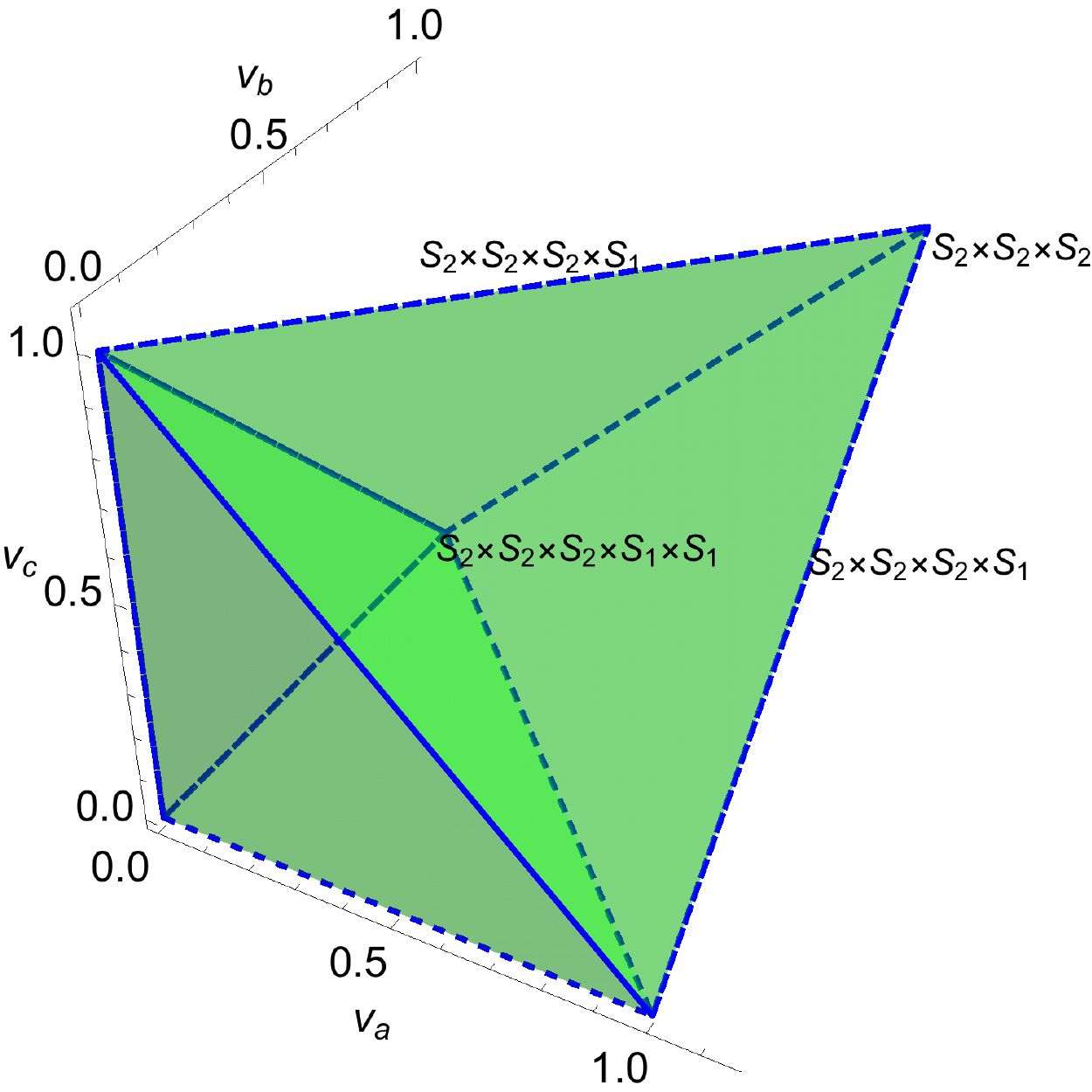}
\caption{The physical region in the ($v_a,v_b,v_c$) graph. It is basically consisted of two tetrahedrons. The local part of the state space is indicated on the outside boundary of the upper tetrahedron.}
\label{fig:vavbvc}
\end{center}
\end{figure}

From the above discussion, we can see that the nonlocal part of the state space basically consists of two tetrahedrons in the ($v_a, v_b, v_c$) graph which are separated by the surface $v_a+v_b+v_c=1$ as shown in fig. \ref{fig:vavbvc}. Inside the tetrahedrons, the state space also includes a sphere (topologically) parametrized by the values $\vbb 2$ and $\vbb 3$ (up to the conjugated pair). The sphere shrinks to a point on the boundary of the combined tetrahedrons.

From the point view of the entanglement, the values of $v_a$, $v_b$ and $v_c$ and therefore the point in the ($v_a,v_b,v_c$) graph represent how much each qubit entangle with the others or equivalently, how much information it shares with the others. On the other hand, the precise position of a state on the inside sphere parametrized by $\vbb 2$ and $\vbb 3$ indicate \emph{how} the qubits entangle with each other. Specifically, the 3-tangle on the sphere can vary between certain range depending on the value of $v_a$, $v_b$ and $v_c$ as we have explained before. This indicates the different points on the same sphere may have different ways of the entanglement or distributing the shared information among the qubits.

To complete the description of the state space, we consider the local degrees of freedom. For the states with all of the three vectors $\vv_a$, $\vv_b$ and $\vv_c$ nonvanishing, there are always local degrees of freedom coming from the directions of the vectors: $S_2\times S_2 \times S_2$. However, we also have the degrees of freedom associated with the local rotations around the spins (i.e., the axial rotations). This part of the degrees of freedom basically coming from the directions of the $\VEC X$-terms. The local degrees of freedom left in that part will depend on the number of the independent directions remaining in the $\VEC X$-terms. The number will change for the different positions in the ($v_a,v_b,v_c$) graph. We can find out the number at a particular position by checking how many expansion probabilities left for the seed state at that point. The result is: 
\begin{itemize}

\item The product state at the tip of the upper tetrahedron: one expansion probability left. There is no $\VEC X$-term so that no local degree of freedom comes from the axial rotations. The local part of the space at this point is $S_2\times S_2 \times S_2$.

\item The bi-separable states at the three lines of the upper tetrahedron that connected to the tip: two expansion probabilities left. There is one $\VEC X$-term and only one local degree of freedom comes from the axial rotations. The local part of the space in this case is $S_2\times S_2 \times S_2 \times S_1$.

\item The states on the surface between two lines of the bi-separable states: three expansion probabilities left. There are three $\VEC X$-terms but only two independent directions. Therefore, there are two local degrees of freedom come from the axial rotations and the local part of the space for this case is $S_2\times S_2 \times S_2 \times S_1 \times S_1$.

\item The general seed states inside of the combined tetrahedrons: four expansion probabilities left. There are six $\VEC X$-terms but only three independent directions. Therefore, there are three local degrees of freedom come from the axial rotations and the local part of the space for this case is $S_2\times S_2 \times S_2 \times S_1 \times S_1 \times S_1$. 

\end{itemize}

We have indicated the local part of the space on the outside boundary of the upper tetrahedron in the fig. \ref{fig:vavbvc}. However, the states on the outside boundary of the \emph{lower} tetrahedron are those states with some of the three vectors vanishing. The local part of the space for those states is not just some trivial product of the three local space but probably some kinds of the quotient space. We will leave the question about the topology of this part of the space open for now.

%%%%%%%%%%%%%%%%%%%%%%%%%%%%%%%%%%%%%%%%%%%%%%%%%%%%
\section{Summary and future directions}

We have provided the method to write down the density operator in MSTA for any pure state of multi-qubit systems. Using this MSTA formulation, we can analyze the quantum problems from a geometric perspective without relying on the conventional matrix methods. We demonstrate some advantages of using the MSTA by some examples including the analysis of the Bell-inequality and the dynamics of two coupled qubits. Lastly, we show how to construct the local unitary invariants in a natural way of the MSTA formulation and analyze the space of the two and three-qubit pure states with the local and non-local degrees of freedom separated.

In this work we mostly focus on the pure states, however, we can also use what we have learned here to study the mixed states.  The density operator of a mixed state is just the probability-weighted sum of the density operators of several pure states. Therefore it should be possible to study the mixed states by using MSTA formulation. By following this approach, it may provide some geometric insight into some entanglement and decoherence problems related to the mixed states.

There are several problems that we have not yet addressed. At first, the formula of $I_6$ in \eqref{I6s} seems very solid and convenient however we have not found a way to derive it. To obtain this formula, we may need to define the corresponding quantity like the concurrence for a mixed state of two qubits in MSTA and follow the procedure in \cite{coffman2000distributed} to derive the 3-tangle. Furthermore, we have explored the basic structures of the state spaces of the two-qubit and three-qubit pure states. However, we have not yet clarified the local parts of the spaces (possibly quotient) of the maximally entangled states of two qubits and the pure states of three qubits with some of the three vectors vanishing. The topology of these spaces seems interesting and may be worth to pursue it further. 

There are several directions we can pursue in the future. At first, it may be possible to define the invariants for the system with an arbitrary finite number of qubits and follow the similar procedure to obtain the nonlocal part of the state space. However, it is unclear that if it is possible to deal with the consistent conditions in general.

Furthermore, we have shown that a time evolution operator can be regarded as a rotor or a series of rotors in MSTA. With this realization, the change of the entanglement in some simple dynamics of two-qubit systems can be seen clearly from a geometric perspective. It will be interesting if we can study the evolution of the entanglement in the dynamics of the system with three (or more than three) qubits in a similar way and show how the states flow in the state space. Additionally, using the MSTA formulation, we may be able to compute the energy levels or the evolution of the Heisenberg spin chain of finite qubits without relying on the Bethe ansatz.

%\footnote{So far, we have been able to compute the exact energy levels for the Heisenberg XXX model up to six qubits.} 

On the practical side, the formulation and computation in MSTA may provide some advantages in dealing with quantum gates and quantum information problem. There are already several works in this direction \cite{havel2000geometric,cafaro2011geometric,alves2010clifford}. Hopefully, we have convinced the readers of the usefulness of the MSTA formulation and more works will follow this approach.

%Lastly, the MSTA formulation can be easily promoted to the relativistic framework as the one shown in \cite{doran2003geometric}. We may be able to use the similar approach to study the multi-qubit system in relativistic regime.

%%%%%%%%%%%%%%%%%%%%%%%%%%%%%%%%%%%%%%%%%%%%%%%%%%%%
%\begin{thebibliography}{99}
%%%%%%%%%%%%%%%%%%%%%%%%%%%%%%%%%%%%%%%%%%%%%%%%%%%%

%\end{thebibliography}
%%%%%%%%%%%%%%%%%%%%%%%%%%%%%%%%%%%%%%%%%%%%%%%%%%

\bibliographystyle{unsrt}
\bibliography{DoGA} 

\begin{thebibliography}{10}

\bibitem{hestenes1966space}
David Hestenes and Anthony~N Lasenby.
\newblock {\em Space-time algebra}, volume~1.
\newblock Springer, 1966.

\bibitem{hestenes1971vectors}
David Hestenes.
\newblock Vectors, spinors, and complex numbers in classical and quantum
  physics.
\newblock {\em American Journal of Physics}, 39(9):1013--1027, 1971.

\bibitem{hestenes1979spin}
David Hestenes.
\newblock Spin and uncertainty in the interpretation of quantum mechanics.
\newblock {\em American Journal of Physics}, 47(5):399--415, 1979.

\bibitem{doran1993states}
Chris Doran, Anthony Lasenby, and Stephen Gull.
\newblock States and operators in the spacetime algebra.
\newblock {\em Foundations of physics}, 23(9):1239--1264, 1993.

\bibitem{doran1996spacetime}
Chris Doran, Anthony Lasenby, Stephen Gull, Shyamal Somaroo, and Anthony
  Challinor.
\newblock Spacetime algebra and electron physics.
\newblock In {\em Advances in imaging and electron physics}, volume~95, pages
  271--386. Elsevier, 1996.

\bibitem{havel2003density}
Timothy~F Havel, CJL Doran, and Suguru Furuta.
\newblock Density operators in the multiparticle spacetime algebra.
\newblock In {\em Proc. Royal. Soc}, 2003.

\bibitem{doran2003geometric}
Chris Doran and Anthony Lasenby.
\newblock {\em Geometric algebra for physicists}.
\newblock Cambridge University Press, 2003.

\bibitem{baylis2012clifford}
William~E Baylis.
\newblock {\em Clifford (Geometric) Algebras: with applications to physics,
  mathematics, and engineering}.
\newblock Springer Science \& Business Media, 2012.

\bibitem{havel2000geometric}
TH~Havel and Chris~JL Doran.
\newblock Geometric algebra in quantum information processing.
\newblock Technical report, 2000.

\bibitem{wie2014bloch}
Chu-Ryang Wie.
\newblock Bloch sphere model for two-qubit pure states.
\newblock {\em arXiv preprint arXiv:1403.8069}, 2014.

\bibitem{wharton2016natural}
KB~Wharton.
\newblock Natural parameterization of two-qubit states.
\newblock {\em arXiv preprint arXiv:1601.04067}, 2016.

\bibitem{sudbery2001local}
Anthony Sudbery.
\newblock On local invariants of pure three-qubit states.
\newblock {\em Journal of Physics A: Mathematical and General}, 34(3):643,
  2001.

\bibitem{acin2001three}
A~Acin, A~Andrianov, E~Jan{\'e}, and Rolf Tarrach.
\newblock Three-qubit pure-state canonical forms.
\newblock {\em Journal of Physics A: Mathematical and General}, 34(35):6725,
  2001.

\bibitem{hill1997entanglement}
Scott Hill and William~K Wootters.
\newblock Entanglement of a pair of quantum bits.
\newblock {\em Physical review letters}, 78(26):5022, 1997.

\bibitem{coffman2000distributed}
Valerie Coffman, Joydip Kundu, and William~K Wootters.
\newblock Distributed entanglement.
\newblock {\em Physical Review A}, 61(5):052306, 2000.

\bibitem{cafaro2011geometric}
Carlo Cafaro and Stefano Mancini.
\newblock A geometric algebra perspective on quantum computational gates and
  universality in quantum computing.
\newblock {\em Advances in Applied Clifford Algebras}, 21(3):493--519, 2011.

\bibitem{alves2010clifford}
Rafael Alves and Carlile Lavor.
\newblock Clifford algebra applied to grover's algorithm.
\newblock {\em Advances in applied Clifford algebras}, 20(3-4):477--488, 2010.

\end{thebibliography}

\end{document}